\documentclass[aps,epsf,twocolumn]{revtex4}
\usepackage{graphicx}
\usepackage{amsmath}
\usepackage{amssymb}
\usepackage{float}
\usepackage{caption}
\usepackage{subcaption}
\usepackage{booktabs}
\usepackage{color}
\usepackage{enumitem}
\usepackage{multirow}

\usepackage[colorlinks=true,
linkcolor=blue,
urlcolor=blue,
citecolor=blue]{hyperref}

\def\prl#1#2#3{{Phys. Rev. Lett.} {\bf #1}, #2 (#3)}

\def\pre#1#2#3{Phys. Rev. E {\bf #1}, #2 (#3)}

\def\epl#1#2#3{{Europhys. Lett.} {\bf #1}, #2 (#3)}

\def\pnas#1#2#3{Proc. Natl. Acad. Sci. (USA) {\bf #1}, #2 (#3)}

\def\physd#1#2#3{Physica D {\bf #1}, #2 (#3)}
\def\physa#1#2#3{Physica A {\bf #1}, #2 (#3)}

\def\prtp#1#2#3{Prog. Theor. Phys. {\bf #1}, #2 (#3)}

\def\ch#1#2#3{Chaos {\bf #1}, #2 (#3)}

\def\natphys#1#2#3{Nat. Phys. {\bf #1}, #2 (#3)}
\def\sc#1#2#3{Science {\bf #1}, #2 (#3)}

\def\biopsych#1#2#3{Biol. Psychiatry {\bf #1}, #2 (#3)}
\def\rpp#1#2#3{Rep. Prog. Phys. {\bf #1}, #2 (#3)}
\def\sciam#1#2#3{Sci. Am. {\bf #1}, #2 (#3)}
\def\apl#1#2#3{Appl. Phys. Lett. {\bf #1}, #2 (#3)}
\def\npcs#1#2#3{Nonlinear Phenom. Complex Syst. {\bf #1}, #2 (#3)}
\def\nonlin#1#2#3{Nonlinearity {\bf #1}, #2 (#3)}
\def\nbr#1#2#3{Neurosci. Biobehav. Rev. {\bf #1}, #2 (#3)}
\def\jpcx#1#2#3{J. Phys. Complex. {\bf #1}, #2 (#3)}
\def\eth#1#2#3{Ethology {\bf #1}, #2 (#3)}
\def\plr#1#2#3{Phys. Life Rev. {\bf #1}, #2 (#3)}
\def\sciadv#1#2#3{Sci. Adv. {\bf #1}, #2 (#3)}
\def\comphys#1#2#3{Commun. Phys. {\bf #1}, #2 (#3)}
\def\njp#1#2#3{New J. Phys. {\bf #1}, #2 (#3)}
\def\pnas#1#2#3{Proc. Natl. Acad. Sci. USA {\bf #1}, #2 (#3)}
\def\sjam#1#2#3{SIAM J. Appl. Math. {\bf #1}, #2 (#3)}
\def\ptrsb#1#2#3{Philos. Trans. R. Soc. B {\bf #1}, #2 (#3)}
\def\ncns#1#2#3{Netw., Comput. Neural Syst. {\bf #1}, #2 (#3)}
\def\pierb#1#2#3{Prog. Electromagn. Res. B {\bf #1}, #2 (#3)}
\def\fhn#1#2#3{Front. Human Neurosci. {\bf #1}, #2 (#3)}
\def\cell#1#2#3{Cell {\bf #1}, #2 (#3)}
\def\bsf#1#2#3{Brain Struct. Funct. {\bf #1}, #2 (#3)}
\def\jneurophysiol#1#2#3{J. Neurophysiol. {\bf #1}, #2 (#3)}
\hypersetup{citecolor=blue}

\graphicspath{{./figures/}}

\begin{document}
	
	\title{Multi-cluster chimeras in phase oscillators with repulsive nonlocal coupling}
	
	\author{Ayushi Saxena}
	\affiliation{Department of Physics, Institute of Science, Banaras Hindu University, Varanasi, Uttar Pradesh 221005, India}
	
	\author{Sangeeta Rani Ujjwal}
	\affiliation{Department of Physics, Institute of Science, Banaras Hindu University, Varanasi, Uttar Pradesh 221005, India}
	\author{Ram Ramaswamy}
\affiliation{Department of Physical Sciences, Indian Institute of Science Education and Research, Berhampur, Odisha 760 003, India}

	
	\begin{abstract}
    
Local repulsive coupling tend to a desynchronize ensembles of globally coupled oscillators, but when the repulsive coupling
is nonlocal, multi-cluster chimeras can result.  In this case, several groups of synchronised oscillators (the so-called clusters) 
are formed, and these coexist with a set of desynchronised oscillators. For phase oscillators on a ring with 
nonlocal piecewise linear repulsive coupling that also involves a phase lag, we find that in the multi-cluster 
chimera state the synchronized clusters are either {\em antiphase} or in {\em splay} with respect to each other, 
namely the $n$ consecutive synchronized clusters differ in phase by $2\pi/n$. This is in contrast to multi-cluster chimeras
that are formed with nonlocal attractive coupling. The synchronized solutions are studied numerically as well as 
analytically and by analysing their stability, we identify the parameter regions where these can be observed.
Our numerical results are validated by dimensional reduction using the Ott-Antonsen analysis. 
		
	\end{abstract}
	
	\maketitle
	
	\section{Introduction}
The natural world is filled with numerous phenomena that involve the interplay of attractive and repulsive interactions that result in fascinating spatio-temporal patterns~\cite{majhi2020}. While many studies have focused on the effects of attractive interactions on the emergent dynamics of the system, it has been noted that the effects of repulsive interactions are not well understood though these interactions play a significant role in deciding the collective dynamics of several real-life systems, for instance, the activity of human brain is the result of excitatory and inhibitory synaptic connections among neurons ~\cite{Neocort2017}. In neuronal systems, excitatory synapses are typically faster than inhibitory synapses, yet inhibition has a greater influence on neuronal firing patterns~\cite {ermentrout, Schwartz}. In social systems, the dynamics of the population can be studied by considering conformists (attractively interacting individuals) and contrarians (repulsively interacting individuals)~\cite{hong2011}. In ecological networks, to understand the coexistence and survival of competitive species, the measures of positive (cooperation) and negative (competition) interactions are used ~\cite{Giron2016}. In the case of coupled quantum oscillators, new kinds of symmetry-breaking transitions can take place due to the presence of attractive-repulsive couplings ~\cite{paul2024}.

 To mimic interactions in real-world systems, hybrid coupling involving attractive and repulsive connections with nonlocal or distance-dependent features is often considered, and these systems show various interesting dynamical states, such as synchronized states ~\cite{strogatz1993}, splay states~\cite{strogatz1993splay,hadley1987}, and chimera states~\cite{kur2002, abrams2004, panaggio2015}. Chimeras are spatiotemporal patterns representing the coexistence of coherent and incoherent behaviour and are relevant to several phenomena, such as unihemispheric sleep~\cite{rattenborg} in some mammals~\cite{haugland2021} and asynchronous eye closure (ASEC) in reptiles~\cite{mathews2006}. Patterns analogous to chimera states are observed in neuronal systems ~\cite{majhi2019, bansal2019, makinwa2023} for instance, abnormal or irregular spatiotemporal patterns may indicate neurological disorders such as Parkinson's disease, epileptic seizures and schizophrenia. Chimeric patterns are observed in diverse settings such as in spiral and scroll waves in heart tissues~\cite{cherry2008}, mechanical oscillators~\cite{martens2013}, pendula networks ~\cite{ebrah2022}, photochemical oscillators~\cite{tinsley2012}, nano electromechanical oscillators ~\cite{matheny2019} and social systems ~\cite{avella2014}. Chimeras were originally reported in networks of identical oscillators with non-local attractive coupling, allowing for both short and long-range connections between oscillators~\cite{kur2002}. Chimeras with multiple coherent clusters termed as multichimeras are reported in coupled oscillators with nonlocal attractive coupling~\cite{omel2012,ujjwal2013}. The combined effect of both attractive and repulsive couplings can lead to chimera death in nonlocally coupled van der Pol oscillators~\cite{sathiyadevi2018}. Also,  sparse long-range inhibitory interactions along with the local excitatory ones can produce stable patterns with non-uniform phase distributions~\cite{ermentrout1994}. Such hybrid coupling finds application in a wide variety of fields, including chemical oscillators, Josephson junction arrays, and neural networks. Therefore, understanding the impact of such hybrid coupling on chimeric patterns can be crucial in understanding various natural phenomena and how information is processed, transmitted, and integrated within networks~\cite{Hutt2014}.

In this work, we study the effects of repulsive nonlocal coupling on the existence and nature of multi-cluster 
chimeras in a system of phase oscillators arranged on a ring. The coupling between the oscillators is distance-dependent, 
piecewise linear, and involves a phase lag. The region(s) of repulsive coupling can exist with the attractive and zero 
coupling in different combinations. In this paper, we investigate the occurrence of different states arising when 
repulsive connections are present in different combinations at lateral and local positions on the ring. Repulsive 
coupling leads to multi-cluster chimeras with phase difference between the synchronised clusters. The phase difference 
can be $\pi$, yielding antiphase chimeras, but in addition, we observe a novel multi-cluster chimera where the phase 
difference between the consecutive clusters is equal: these are {\em splay} chimeras, which have not been 
reported earlier to the best of our knowledge. We further study the effect of  variation of the phase lag parameter, $\alpha$
on the complex dynamics as well as other modifications of the nonlocal coupling function. 

This paper is organised as follows. In Sec.~\ref{sec:model} the model is described and the emergent states for different 
combinations of attractive and repulsive couplings at local and lateral positions are discussed. The synchronized solutions 
and their stability with the phase lag parameter are presented in Sec.~\ref{sec:sync-sol}. Phase switching in two cluster 
chimeras and the effects of repulsive connections on synchronized clusters are examined in Sec.~\ref{sec:2c-switch}. 
The effects of repulsive interactions on multi-cluster chimeras are discussed in Sec.~\ref{sec:multi}. Numerical results 
are verified using the Ott-Antonsen analysis presented in Sec.~\ref{sec:dim-reduction} and this is followed by 
a summary and discussion in Sec.~\ref{sec:summary}.  
          
\section{Model description and emergent states} \label{sec:model}

We consider the Sakaguchi-Kuramoto model~\cite{sakaguchi1986} of identical phase oscillators arranged on a ring, governed by the dynamical equation
\begin{equation}
	\frac{\partial \phi(x,t)}{\partial t} = \omega - \int_{-\pi}^{\pi}G(x-x')\sin(\phi(x,t) - \phi(x',t) + \alpha)dx',
	\label{eq:eq1}
\end{equation}
where $\phi(x,t)$ is the phase of the oscillator at position $x$ and time $t$. The phase lag parameter, $\alpha$ is 
constrained to the interval [0, $\pi$/2]. The intrinsic frequency, $\omega$ is the same for all oscillators and 
taken to be equal to zero. The non-local coupling is introduced via the distance-dependent kernel, $G(x - x')$ 
which can take positive, negative or zero values and is considered to be piecewise linear~\cite{ujjwal2013},
symmetric and normalized to unity. 
For a finite number of oscillators, Eq.\eqref{eq:eq1} can be discretised, with the position of $i^{th}$ oscillator
being $x_{i}$ = -$\pi + (2\pi i/N$),  $i = 1, 2, 3,\ldots, N$ and periodic boundary conditions are applied.  

Applying the self-consistency argument for globally coupled oscillators as proposed by Kuramoto~\cite{kur2002}, 
the system dynamics can be described by $\Phi(x,t)$ = $\phi(x,t)$ - $\zeta t$, where $\zeta$ represents the 
angular frequency of a rotating frame, and $\Phi$ is the phase of an oscillator relative to this frame. 
The complex order parameter, denoted as $R e^{i\Theta}$ is dependent on both space and time and is expressed as
\begin{equation}
	R(x,t)e^{i\Theta(x,t)} =  \int_{-\pi}^{\pi}G(x-x')e^{i\Phi(x',t)}dx'.
	\label{eq:eq2}
\end{equation}
Therefore Eq.\eqref{eq:eq1} in terms of the components of the order parameter becomes
\begin{equation}
	\frac{\partial \Phi(x,t) }{\partial t} = \omega - \zeta - R\sin(\Phi(x,t) -\Theta +\alpha),
	\label{eq:eq3}
\end{equation}
where the phases of the oscillators are decoupled, but the oscillators still interact through the 
coherence parameter $R$ and average phase $\Theta$.
	

\begin{figure}[h!]
	\centering
	\includegraphics[width=1.0\linewidth]{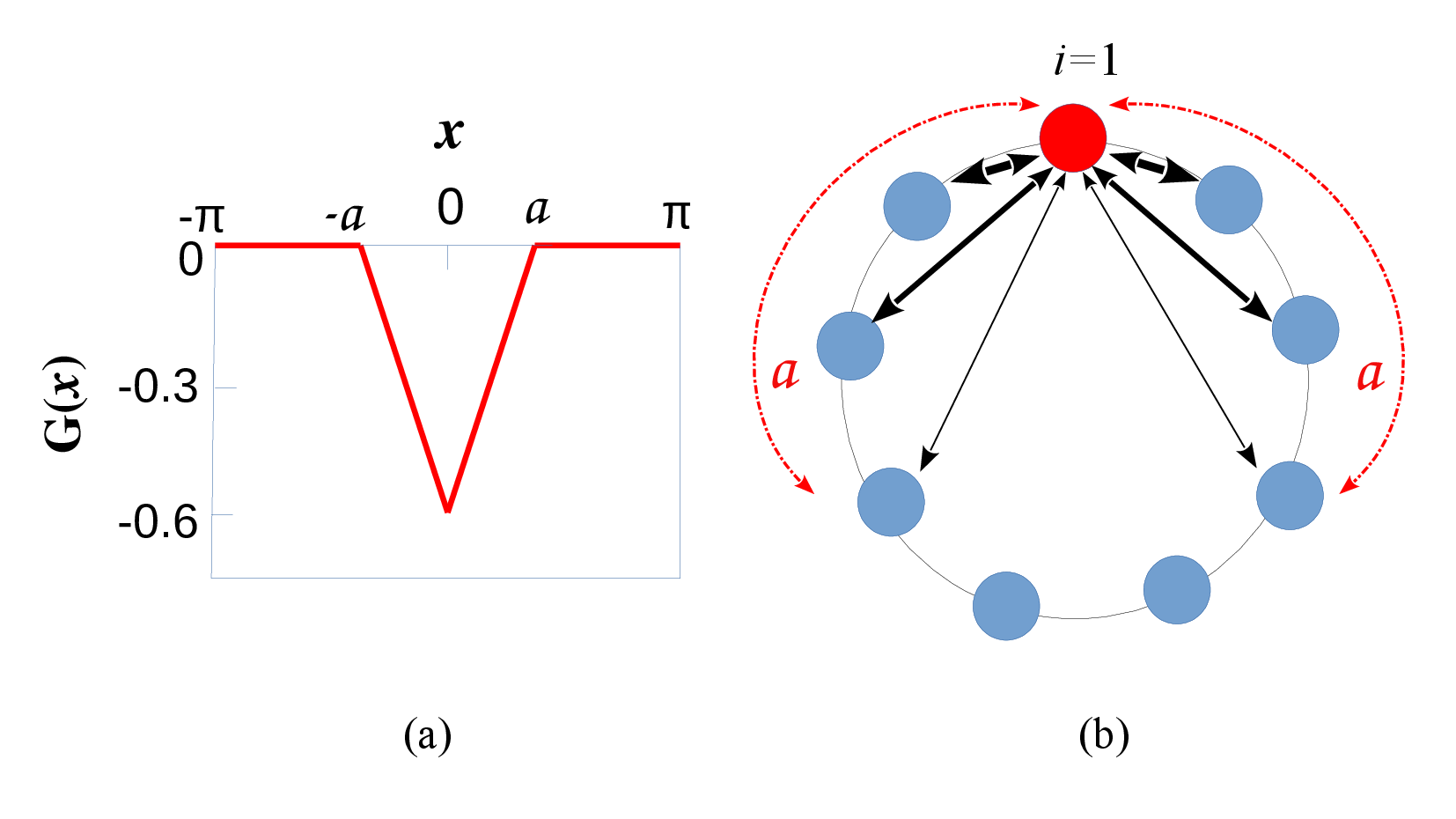}
	\captionsetup{skip=0pt}
	\caption{Schematic diagram showing the nonlocal piecewise linear coupling depicting local repulsive and lateral neutral interactions with finite coupling range `$a$' symmetric about $x=0$. (a) Functional form of $G(x)$ with number of coupling segments, $n_s$=2 and (b) coupling between oscillators on the ring, where the thickness of the arrows indicates the magnitude of repulsive coupling with respect to the first($i=1$) oscillator (shown with a red circle).}
	\label{fig:Fig1}
	
\end{figure}

We study the dynamics of the system resulting from different forms of nonlocal coupling: purely attractive, purely repulsive and a mix of
different ranges of attractive, repulsive and null interactions. The chimera states for different combinations of 
positive and zero coupling regions have been reported earlier~\cite{ujjwal2013}. In this work our focus is on 
chimeras when the coupling is finite range repulsive, an example of which is shown in Fig.~\ref{fig:Fig1}. 
An analogy to this kind of coupling can be found in the connections between cortical neurons and their neighbours, 
where the coupling kernel resembles ``inverted Mexican-hat" like function, comprising short-range inhibition and long-range excitation or neutral connections~\cite{hutt2003}.
\begin{figure}[h!]
	\centering
	\includegraphics[width=1.0\linewidth]{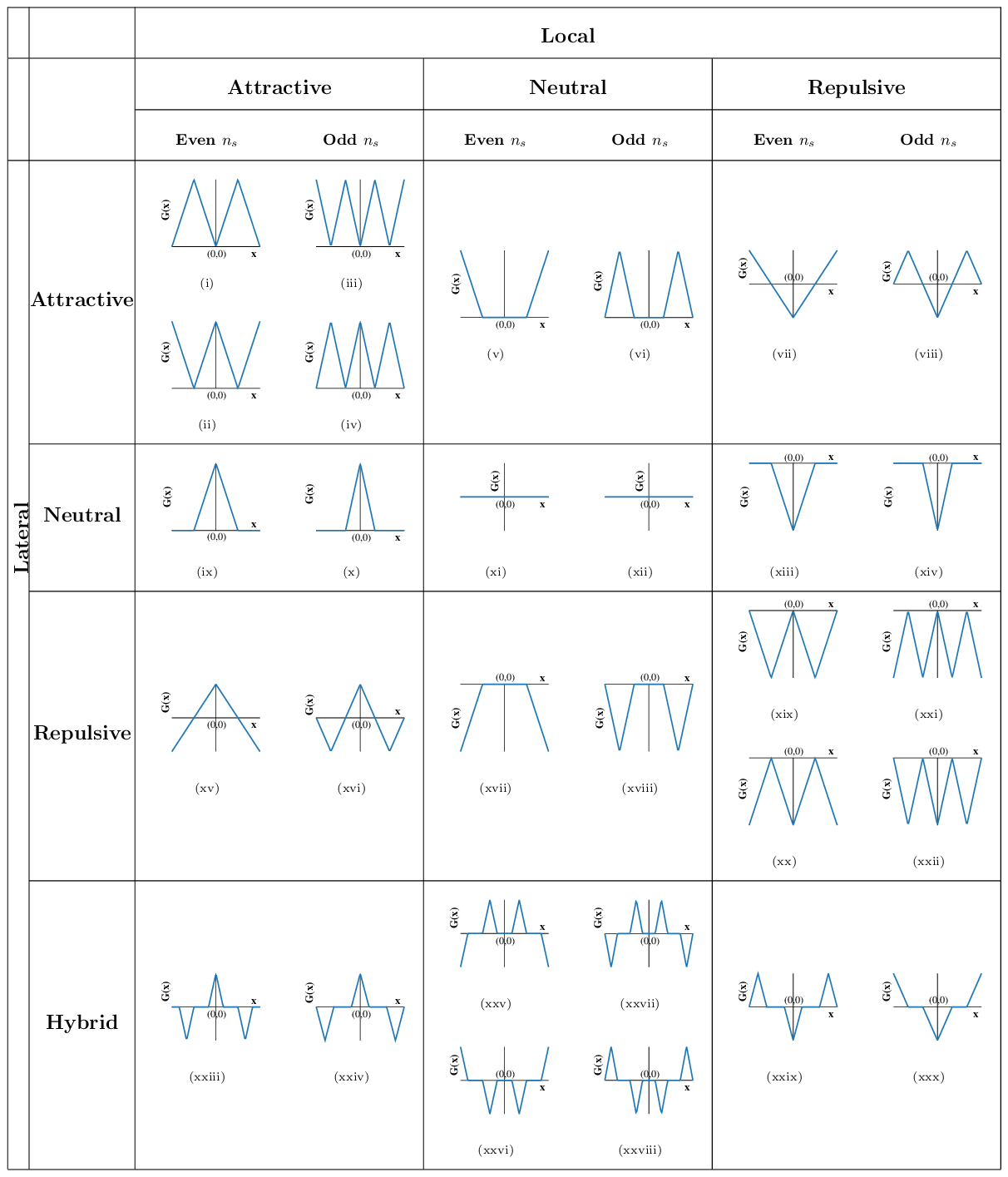}
	\captionsetup{skip=0pt}
	\caption{Figure showing various forms of nonlocal piecewise linear coupling function $G(x)$. The coupling kernel, $G(x)$ is categorized based on the nature of coupling present locally and laterally. Further, $G(x)$ is classified by looking at the odd-even parity of the number of coupling segments in the positive $x$ direction, $n_s$. }
	\label{fig:Fig2}
	
\end{figure}

\begin{figure}[htbt!]
	\centering
	\includegraphics[width=\linewidth]{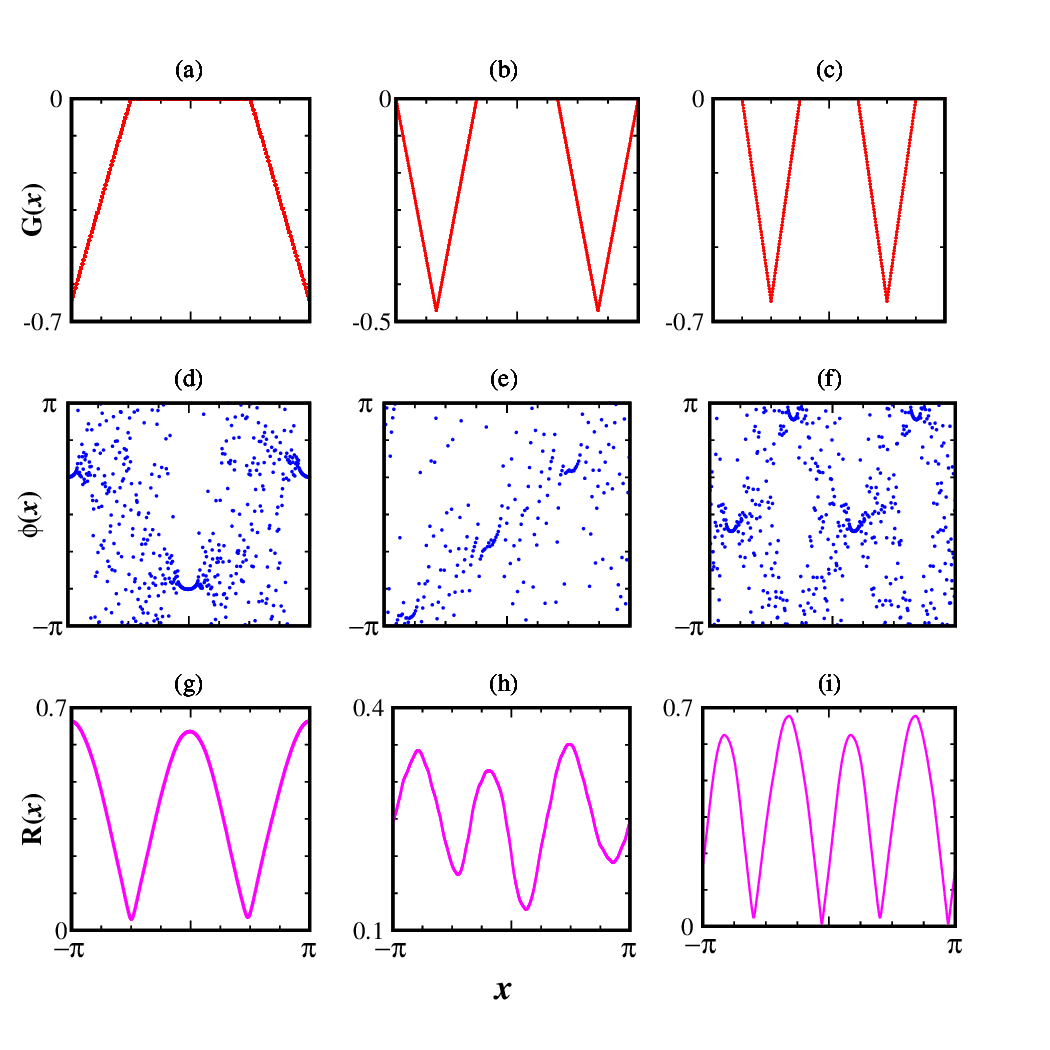}
	\caption{Figure showing asymptotic phase profile $\phi(x)$ and the corresponding local order parameter $R(x)$ for $G(x)$ with lateral repulsive coupling with no local connections. The functional form of $G(x)$ is shown in the top panel ((a)-(c)). $\phi(x)$ and $R(x)$ for $\alpha=1.52$ are shown in the second and third rows respectively. $G(x)$ with even $n_s$ gives rise to antiphase chimeras ($APC_{n_s}$) ([(a),(d),(g)] for $n_s$=2 and [(c),(f),(i)] for $n_s$=4) whereas $G(x)$ with odd $n_s$ asymptotes to a splay chimera ($SC_{n_s}$) ([(b),(e),(h)] for $n_s$=3). For simulations, the initial phases are taken uniformly from $ (-\pi, \pi)$. The system's equation Eq.\eqref{eq:eq1} is integrated using the Runge-Kutta fourth-order scheme with step size $h$ = 0.01. The phases and order parameters are plotted after discarding 5 $\times 10^5$ transients for $N$=1024.}
	\label{fig:Fig3}
\end{figure}

The effects of repulsive nonlocal coupling on the emergent dynamical states are studied by numerically simulating Eq.\eqref{eq:eq1} and then analysing the asymptotic states. The interplay of the coupling function and the phase lag parameter $\alpha$ gives rise to various states. Low values of $\alpha$ favour synchronized states whereas $\alpha$ close to $\pi/2$ gives rise to chimeras. We fix $\alpha=1.52$ and study the effect of changing the form of $G(x)$ on the emergent states. We extensively investigate different combinations of repulsive, attractive and zero coupling regions in $G(x)$ and the resulting states. From our earlier study~\cite{ujjwal2013} we know that the number of segments of finite and zero coupling decides the number of clusters in emerging chimeras. We denote the number of segments of distinct coupling regions in the positive $x$ direction by $n_s$. Different forms of $G(x)$ are classified according to the nature of the local and lateral coupling and whether $n_s$ is odd or even. Some of the coupling kernels $G(x)$ from different categories are shown in Fig.~\ref{fig:Fig2}. When $n_s$ is small, say 2 or 3, one can clearly distinguish between local and lateral couplings (Fig.~\ref{fig:Fig2}(i)-(xxii)), for instance, Fig.~\ref{fig:Fig2}(vii) depicts a locally repulsive and laterally attractive coupling with $n_s$=2. As $n_s$ becomes large and $G(x)$ takes a more complex form, it is difficult to classify the lateral coupling as purely repulsive, attractive or zero. Therefore in these cases we call the lateral coupling `hybrid' (Fig.~\ref{fig:Fig2}(xxiii)-(xxx)).
 However, if the coupling is repulsive laterally with no local connectivity, the system evolves to an antiphase chimera ($APC_{n_s}$) for even $n_s$ (Fig.~\ref{fig:Fig3}(a),(d),(g) for $n_s$=2 and Fig.~\ref{fig:Fig3}(c),(f),(i) for $n_s$=4) and to a splay chimera ($SC_3$) for odd $n_s$ (Fig.~\ref{fig:Fig3}(b),(e),(h) for $n_s$=3). In a splay chimera with $n$ number of synchronized clusters denoted by $SC_n$, the consecutive synchronized clusters have an equal phase difference equal to $2\pi/n$.

The states emerging from the coupling kernel $G(x)$ shown in Fig.~\ref{fig:Fig2} are summarised in Table~\ref{Table1}. From the investigation of different $G(x)$, we can infer that a local repulsive coupling favors a desynchronized state most of the time, irrespective of the nature of lateral coupling. However, for some cases, one can also observe antiphase chimeras ($APC_{2n_s}$) (Fig.~\ref{fig:Fig2}(viii),(xix), (xxi)), which is a bit counterintuitive. We also observe that repulsive connections always induce a phase shift between the synchronized clusters in multi-cluster chimeras. Therefore with repulsive coupling, we either get a DS or a chimera states with antiphase clusters ($APC_n$) or a splay chimera ($SC_n$), but never chimeras with in-phase clusters ($IPC_n$). $APC_n$ is possible only for even $n$, whereas ($SC_n$) can have both even or odd numbers of clusters, $n$. There is no specific relation between the odd-even parity of $n_s$ and the nature of the emergent multi-cluster chimera. To get in-phase chimeras ($IPC_n$) purely attractive coupling (with or without zero coupling regions) is required.

\begin{table}[h]
\centering
\renewcommand{\arraystretch}{2.8}
\setlength{\tabcolsep}{0.8pt}
\resizebox{\columnwidth}{!}{
\begin{tabular}{|c|c|cc|cc|cc|}
\hline

&  & \multicolumn{6}{c|}{\textbf{ Local}} \\ \hline

\multirow{5}{*}{\rotatebox{90}{\textbf{ Lateral}}}
&
& \multicolumn{2}{c|}{\textbf{ Attractive}}
& \multicolumn{2}{c|}{\textbf{ Neutral}}
& \multicolumn{2}{c|}{\textbf{ Repulsive}} \\
\cline{3-8}

&
& \textbf{Even $n_s$} & \textbf{Odd $n_s$}
& \textbf{Even $n_s$} & \textbf{Odd $n_s$}
& \textbf{Even $n_s$} & \textbf{Odd $n_s$} \\
\cline{2-8}

& \textbf{ Attractive}
& \shortstack{FS \\[1pt] $IPC_{n_s}$}
& \shortstack{FS \\[1pt] $IPC_{n_s}$}
& $IPC_{n_s}$
& $IPC_{n_s}$
& $DS$
& $APC_{2n_s}$ \\\cline{2-8}

& \textbf{ Neutral}
& $APC_{n_s}$
& $APC_{(n_s+1)}$
& $DS$
& $DS$
& $DS$
& $DS$ \\\cline{2-8}

& \textbf{ Repulsive}
& $APC_{n_s}$
& $SC_{n_s}$
& $APC_{n_s}$
& $SC_{n_s}$
& \shortstack{$APC_{2n_s}$ \\[0.2pt] $DS$}
& \shortstack{$APC_{2n_s}$ \\[0.2pt] $DS$} \\\cline{2-8}

& \textbf{ Hybrid}
& $DS$
& $SC_{n_s}$
& \shortstack{$SC_{n_s}$ \\[0.2pt] $SC_{n_s}$}
& \shortstack{$SC_{n_s}$ \\[0.2pt] $DS$}
& $DS$
& $DS$ \\ \hline

\end{tabular}
}
      	
	\caption{ Table summarizing the asymptotic states at $\alpha$ = 1.52 for different coupling functions, $G(x)$ shown in Fig.~\ref{fig:Fig2}. The states observed are in-phase chimera ($IPC_{n_s}$), antiphase chimera ($APC_{n_s}$), splay chimera ($SC_{n_s}$), completely synchronized state ($FS$), and desynchronized states ($DS$), where the subscript denotes the number of synchronized clusters in a chimera state.}
	\label{Table1}
\end{table}

\section{Synchronized solutions}\label{sec:sync-sol}

For low values of $\alpha$ splay states ($SS$) are observed when the coupling is finite range repulsive with positive or null local coupling regions. These results are in line with earlier observations where splay states are reported in networks with inhibitory connectivity~\cite{zillmer2006}. The synchronized solutions and its stability can be computed analytically. We rewrite Eq.\eqref{eq:eq1} in terms of the oscillator indices. Therefore, the phase of the $i^{th}$ oscillator will evolve according to the following equation:

\begin{equation}
	\frac{\partial \phi_i}{\partial t} =  \sum_{j=1}^{N}G_{ij} \sin(\phi_j - \phi_i - \alpha).
	\label{eq:eq4}
\end{equation} 	

Expanding the coupling term gives

\begin{equation}
	\frac{\partial \phi_i}{\partial t} =  \sum_{j=1}^{N}G_{ij} [ \cos\alpha \sin(\phi_j - \phi_i ) - \sin\alpha \cos(\phi_j - \phi_i )].
	\label{eq:eq5}
\end{equation} 

Assuming a solution for the synchronised state as

\begin{equation}
	\phi_i =  \Omega t + \delta_i,
	\label{eq:eq6}
\end{equation} 

where $\Omega$ is the synchronized frequency of the oscillators and $\delta_i$ is the phase of the $i^{th}$ oscillator with respect to the first oscillator in the synchronized state. Putting this solution in Eq.\eqref{eq:eq5} gives

\begin{equation}
	\Omega =  \sum_{j=1}^{N}G_{ij} [ \cos\alpha \sin(\delta_j - \delta_i ) - \sin\alpha \cos(\delta_j - \delta_i )].
	\label{eq:eq7}
\end{equation} 

Taking the average over all the oscillators, the first term in Eq.\eqref{eq:eq7} vanishes and the equation for the collective frequency, $\Omega$ becomes

\begin{equation}
	\Omega =  - \frac{\sin\alpha}{N}\sum_{i=1}^{N}\sum_{j=1}^{N}G_{ij}   \cos(\delta_j - \delta_i ).
	\label{eq:eq8}
\end{equation} 

Introducing the notation $\Delta_{ij}$ = $\delta_j$ - $\delta_i$, Eq.\eqref{eq:eq8} reads

\begin{equation}
	\Omega =  - \frac{\sin\alpha}{N}\sum_{i,j}^{}G_{ij}   \cos\Delta_{ij}.
	\label{eq:eq9}
\end{equation} 	
We can obtain the estimate of $\Omega$ by expanding the right-hand side of Eq.\eqref{eq:eq9}

\begin{equation}
	\Omega =  - \frac{\sin\alpha}{N}\sum_{i,j}^{}G_{ij}[1-\frac{\Delta_{ij}^2}{2}+ \ldots],
	\label{eq:eq10}
\end{equation} 

\begin{equation}
	\Omega =  - \frac{\sin\alpha}{N}\sum_{i,j}^{}G_{ij}  + \frac{\sin\alpha}{2N}\sum_{i,j}^{}G_{ij}\Delta_{ij}^2 - \ldots.
	\label{eq:eq11}
\end{equation} 

The mean-field approximation corresponds to the collective frequency,

\begin{equation}
	\Omega =  - \sin\alpha \langle G \rangle + \frac{\sin\alpha}{2} \langle G\Delta^2\rangle - \ldots,
	\label{eq:eq12}
\end{equation} 

where $\langle . \rangle$ denotes the average over all the oscillators, such that

\begin{equation}
	\langle G \rangle =  \frac{1}{N} \sum_{i,j}^{} G_{ij}, 
	\langle G\Delta^2 \rangle =  \frac{1}{N} \sum_{i,j}^{} G_{ij}\Delta_{ij}^2, \ldots.
	\label{eq:eq13}
\end{equation}

\begin{figure}[htbt!]
	\centering
	\includegraphics[width=\linewidth]{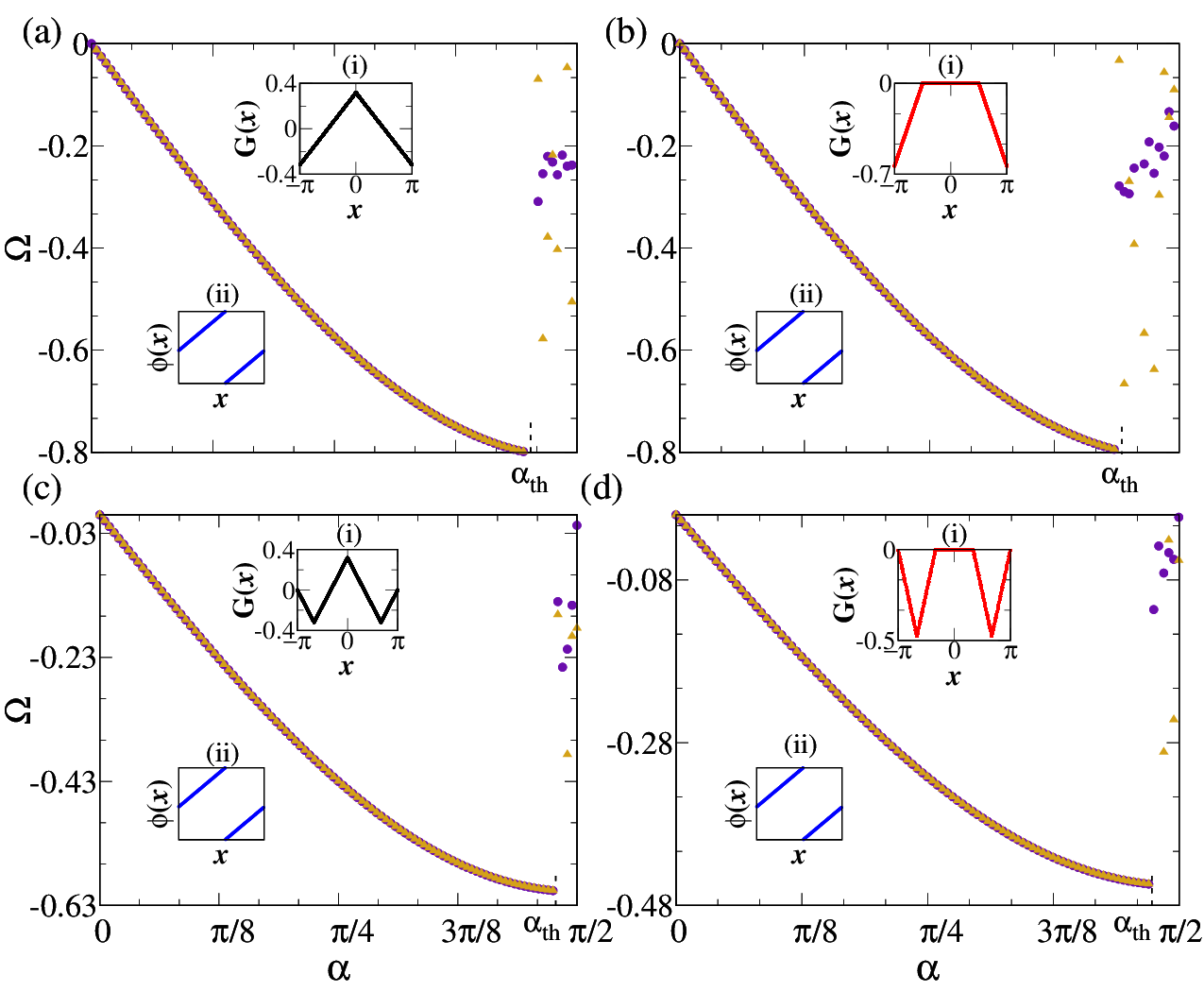}	
	\caption{ Synchronized frequency $\Omega$ is shown with the variation of $\alpha$ for different coupling kernel $G(x)$. Purple circles represent the analytical solutions obtained from Eq.~\eqref{eq:eq9} and the brown triangles are the numerically obtained frequency computed by integrating Eq.~\eqref{eq:eq1}. In each subfigure[(a)-(d)], the inset shows (i)the coupling kernel, $G(x)$ and corresponding phase profile(s) $\phi(x)$ of the synchronized state in (ii). The plots are obtained for $N$ = 128, discarding $3 \times10^5$ transients.}
	\label{fig:Fig4}
\end{figure}

The first term in Eq.~\eqref{eq:eq13} captures the network topology, while the latter terms depend on phase differences as well. The collective frequency $\Omega$ obtained analytically (purple circles) from Eq.~\eqref{eq:eq9} is plotted with the variation of phase lag parameter, $\alpha$ in Fig.~\ref{fig:Fig4}. For comparision numerical solutions for different coupling schemes obtained from integrating Eq.~\eqref{eq:eq1} are also shown with brown triangles. The plots are presented for $G(x)$ where repulsive coupling is present in combination with attractive or null coupling regions. From Fig.~\ref{fig:Fig4} we observe that the analytical and numerical solutions match well in the $\alpha$ range where synchronized solutions (or $SS$) exist. The form of $G(x)$ and the phase profile of the system corresponding to the synchronized state are shown in insets. The analytical solutions begin to deviate from the numerical results as $\alpha$ increases beyond a threshold $\alpha_{th}$ at which synchronized solutions cease to exist. Chimera states exist for $\alpha> \alpha_{th}$ indicated by randomly scattered values of $\Omega$ in Fig.~\ref{fig:Fig4}.

\begin{figure}[htbt!]
	\centering
	\includegraphics[width=0.9\linewidth]{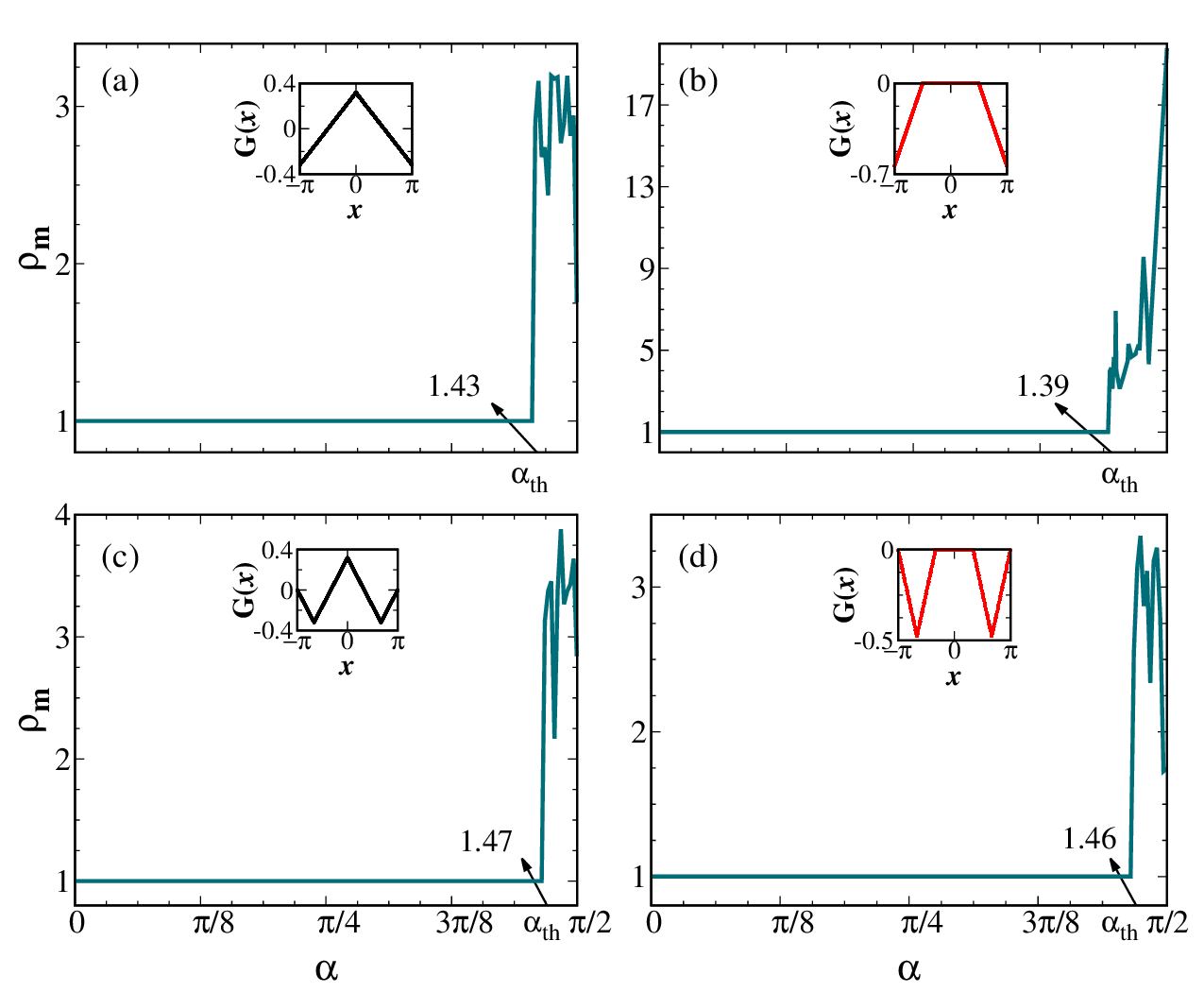}
	\caption{ The largest Floquet multiplier $\rho_{m}$ with $\alpha$ for different coupling kernels, $G(x)$ (shown in the insets) corresponding to the states presented in Fig.~\ref{fig:Fig4}.}
	\label{fig:Fig5}
\end{figure}

\begin{figure}[htbt!]
	\centering	
	\includegraphics[width=0.79\linewidth]{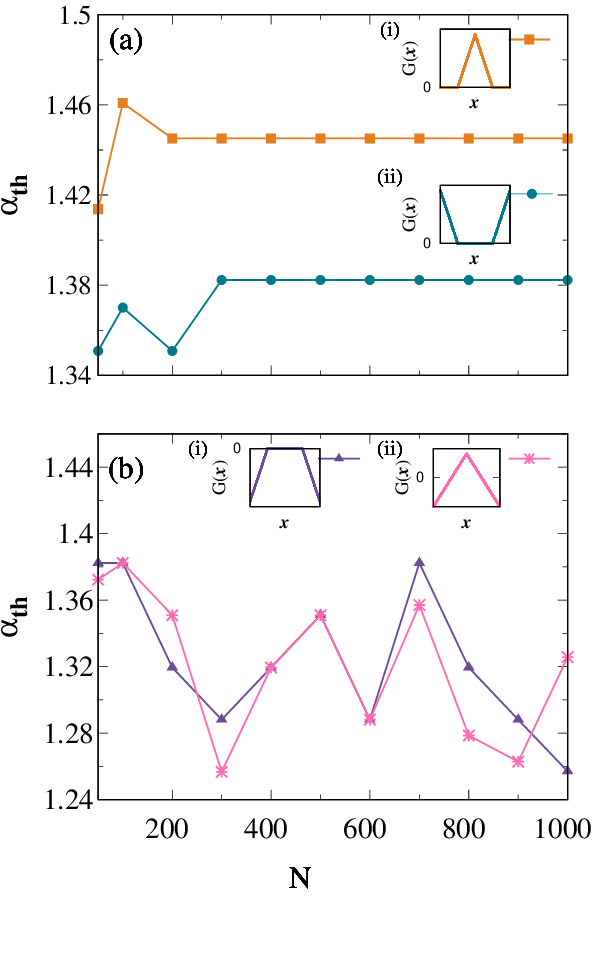}
	\caption{ Variation of $\alpha_{th}$ with system size $N$ for different forms of $G(x)$ with (a) attractive and (b) repulsive coupling regions. The coupling functions $G(x)$ are shown in the insets [(i) and (ii)]. }
	\label{fig:Fig6}
\end{figure}

The stability of synchonized solutions ($SS$ in the present case) is tested by computing Floquet multipliers~\cite{Jordan} of the system that describe the behaviour of perturbations around the periodic solutions with time period T = 2$\pi$/$\Omega$. For a stable synchronized solution with frequency $\Omega$, the largest Floquet multiplier, $\rho_{m}$ is close to unity while $\rho_{m} > 1$ indicates the loss of stability of the synchronized state as $\alpha$ increases (see Fig.~\ref{fig:Fig5}). From Fig.~\ref{fig:Fig5} it can be seen that $\alpha_{th}$ varies across coupling configurations. The $\alpha_{th}$ also changes with the system size $N$. For $G(x)$ containing attractive coupling (locally or laterally) without repulsive coupling, $\alpha_{th}$ saturates to a particular value as $N$ increases (Fig.~\ref{fig:Fig6}(a)). In contrast, for $G(x)$ with repulsive coupling regions, no such trend is observed and $\alpha_{th}$ fluctuates with $N$ (see Fig.~\ref{fig:Fig6}(b)). 

\section{Switching from in-phase to antiphase chimeras }\label{sec:2c-switch}

\begin{figure}[htbt!]
	\centering
	\includegraphics[width=0.7\linewidth]{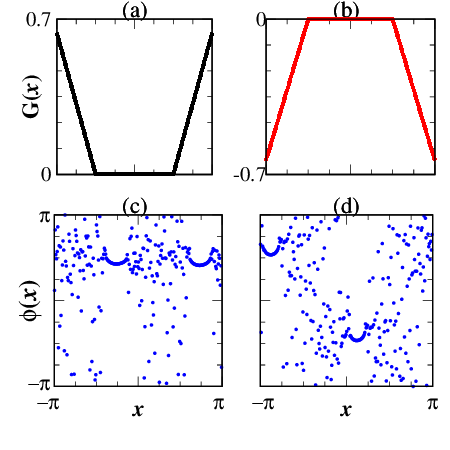}
	\captionsetup{skip=0pt}
	\caption{ Figure showing the switching of two cluster in-phase chimera ($IPC_2$) to a two cluster antiphase chimera ($APC_2$) when lateral attractive coupling is changed to repulsive for $N$ = 1024 and $\alpha$ = 1.52.}
	\label{fig:Fig7}
\end{figure}

In earlier work~\cite{ujjwal2013} a recipe was provided for designing $G(x)$ using piecewise attractive coupling regions
combined with regions of null coupling in order 
to obtain multi-cluster chimeras with specific arrangements of in-phase and anti-phase clusters. An interesting question 
to investigate is how these multi-cluster chimeras change when the attractive regions in $G(x)$ become repulsive. Since the 
coupling is piecewise linear, this can be done in a straightforward manner through a sign-change. For the 
single cluster chimera (C$_1$), when the attractive coupling is made repulsive, the resulting state is completely 
desynchronized. Lateral attractive and local zero coupling result in a two-cluster in-phase chimera ($IPC_2$) 
(Fig.~\ref{fig:Fig7}(a),(c)). When this lateral attractive coupling is made repulsive, we observe a switch 
from in-phase two-cluster chimera ($IPC_2$) to a two-cluster antiphase chimera ($APC_2$). This observation 
of transition from in-phase to antiphase chimeras in coupled oscillators is novel.

\begin{figure}[htbt!]
	\centering
	\includegraphics[width=0.9\linewidth]{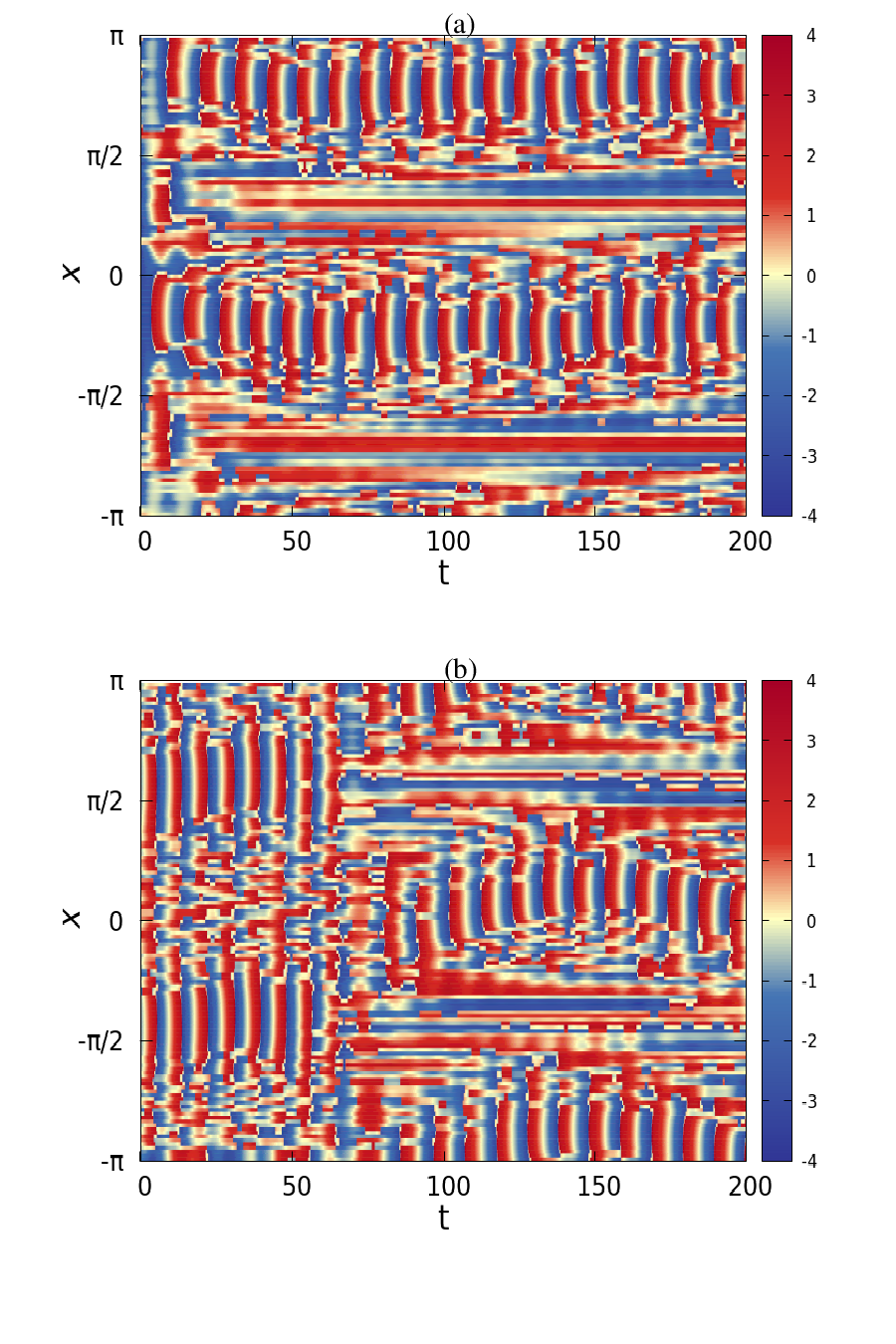}
	\captionsetup{skip=0pt}
	\caption{ Space-time plots showing the emergence of two-cluster antiphase chimera ($APC_2$) from (a) a single-cluster chimera ($C_1$) and (b) two-cluster in-phase chimera ($IPC_2$). The colour bar represents the phase of the oscillators.}
	\label{fig:Fig8}
\end{figure}

We check the robustness of the antiphase state arising from lateral repulsive coupling (see Fig.~\ref{fig:Fig7}(b)) with respect to the choice of initial phases. For this, we start with a single-cluster chimera (C$_1$) as an initial condition and track the emergence of the antiphase chimera state (\href{https://github.com/ayushisaxena0019/gif/blob/main/singleC.gif}{movie-1}). Similarly, starting with a two-cluster in-phase chimera, the system again evolves to a two-cluster antiphase chimera(\href{https://github.com/ayushisaxena0019/gif/blob/main/antiC.gif}{movie-2}). Furthermore, random initial conditions consistently lead to the antiphase chimera. The space-time plots for the emergence of $APC_2$ from $C_1$ and $IPC_2$ as initial conditions are shown in Fig.~\ref{fig:Fig8}. Interestingly, the two antiphase clusters in $APC_2$ move with time; however, the rate of their movement depends upon the initial condition from which they have evolved.  

\begin{figure}[htbt!]
	\centering
	\includegraphics[width=0.81\linewidth]{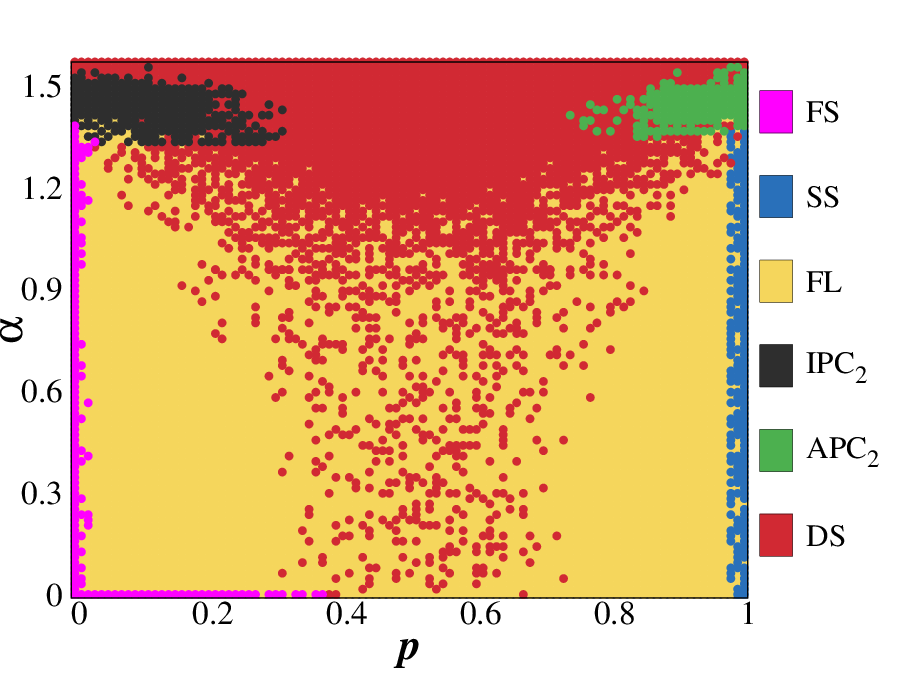}
	\caption{ Figure showing the existence of various dynamical states in $\alpha$-$p$ space. Black dots represent two-cluster in-phase chimera ($IPC_2$), green dots are for two-cluster antiphase chimeras ($APC_2$), red dots show desynchronized state ($DS$), yellow dots are for the frequency locked scattered phase ($FL$), blue dots represent the splay state ($SS$), and the pink dots depict the fully synchronized state ($FS$). The states are computed for $N$ = 128 in interval $\delta \alpha$ = 0.0157 and $\delta p$ = 0.01. }
	\label{fig:Fig9}
\end{figure}

\begin{figure}[htbt!]
	\centering	
	\includegraphics[width=1.0\linewidth]{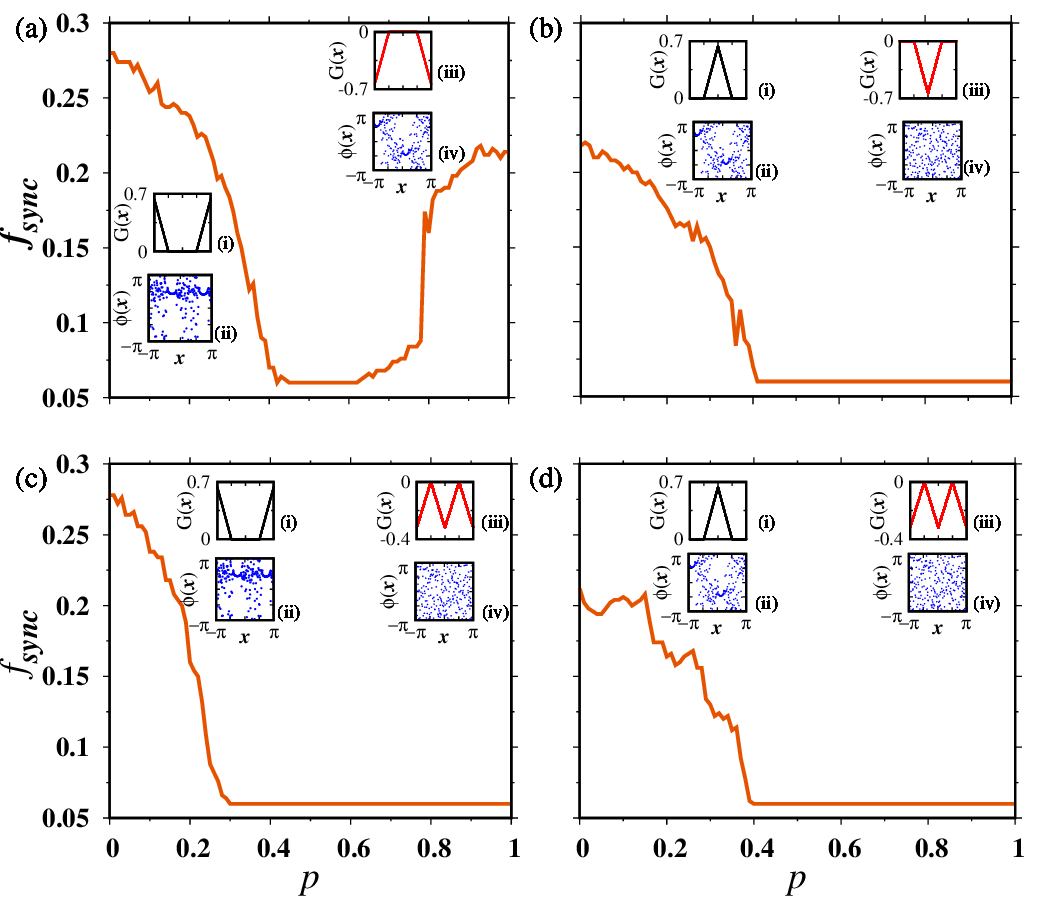}	
	\caption{ The fraction of oscillators forming the synchronized cluster(s) $f_{sync}$ with probability $p$ is shown when the repulsive coupling is applied to the attractive coupling region only (upper panel) and when repulsive coupling is applied to the entire coupling range (including zero coupling regions) (lower panel). The coupling kernel $G(x)$ gives rise to $IPC_2$ [(a),(c)] and $APC_2$ [(b)(d)] at $p=0$. The coupling function, $G(x)$ and the corresponding phase state, $\phi(x)$ for $p$=0 [(i)(ii)] and $p$=1[(iii)(iv)] are shown in the insets. The plots shown are obtained by averaging over 100 realizations for $N$=512.}
	\label{fig:Fig10}
\end{figure}

\begin{figure}[htbt!]
	\centering	
	\includegraphics[width=0.8\linewidth]{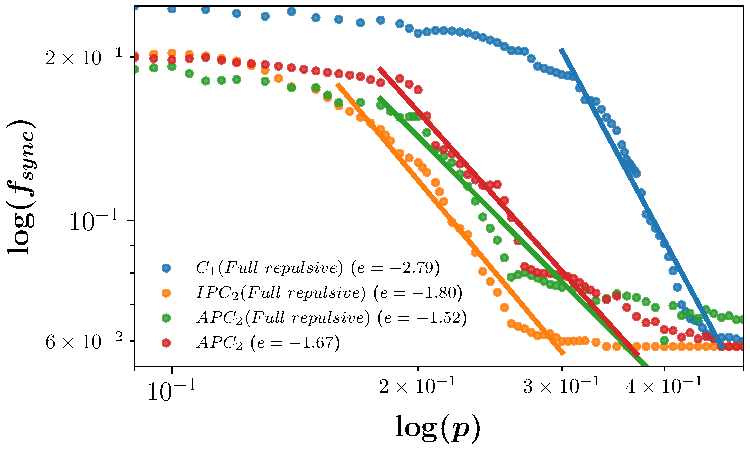}	
	\caption{ Log-log plot of $f_{sync}$ with probability $p$ showing power law in the collapse of chimeras to desynchronized state for $C_1$, $IPC_2$, and $APC_2$.}
	\label{fig:Fig11}
\end{figure}

To probe further into the switching of in-phase chimera to antiphase chimera, we make the attractive coupling repulsive in a probabilistic manner by introducing repulsive connections with probability $p$ such that $p$=0 implies finite-range purely attractive coupling and $p$=1 means purely repulsive coupling. In this context, we consider two cases: first, the repulsive coupling is introduced only in the finite range attractive coupling region, keeping the neutral coupling region unaltered, and in the later case the repulsive coupling is introduced in the entire coupling range (including the zero coupling region). For the first case occurrence of different dynamical states in the $\alpha- p$ parameter space is shown in Fig.~\ref*{fig:Fig9}. Different colours represent different states of the system. For small values of $\alpha$ and $p$, the coupling between the oscillators is mostly attractive, which promotes coherence in the system. In this regime, the system exhibits fully synchronized states ($FS$) shown with pink color. The $FS$ state appears only in a narrow region of the parameter space where $p$ is close to 0, and the dominance of attractive coupling enables the oscillators to achieve complete synchronization. As the phase lag $\alpha$ increases at low $p$, the interaction between oscillators becomes frustrated. This frustration leads to the formation of two-cluster in-phase chimeras ($IPC_2$) (black region). For higher $\alpha$ and moderate $p$, the combined effect of phase lag and repulsive coupling suppresses the phase coherence and tends the system towards desynchronized state ($DS$) shown in red in Fig.~\ref{fig:Fig9} . At $p$ close to 1 and higher $\alpha$ values, the system evolves to a two-cluster antiphase chimera ($APC_2$) in which the two synchronized clusters are out of phase (green region). At small $\alpha$, the system exhibits splay states ($SS$) when the repulsive connections dominate the coupling ($p\approx 1$). For a large region in $\alpha- p$ space, frequency locked states ($FL$) are observed where all the oscillators have equal freqency but their phases do not show any specific pattern (shown in yellow in Fig.~\ref{fig:Fig9}). For intermediate $p$ values frequency locked ($FL$) states coexist with the desynchronized state ($DS$). For lower values of $p$, the dominance of attractive connections allows the oscillators to achieve full synchrony. At intermediate values of $p$, the interplay between attractive and repulsive couplings introduces frustration in the phase dynamics, leading to irregular phase distribution while maintaining frequency locking, hence giving rise to $FL$ states. As $p$ increases further, the repulsive interactions dominate, inducing phase differences between the oscillators, and the oscillators organise into a splay state ($SS$). This seems to suggest that purely attractive coupling favours in-phase states giving rise to complete synchrony ($FS$) for low $\alpha$ and $IPC_2$ at higher $\alpha$ while repulsive connections induce phase difference among the synchronized oscillators resulting in splay states ($SS$) for low $\alpha$ and $APC_2$ at higher $\alpha$ values. The competition between phase lag, attractive, and repulsive coupling can lead to multistability between $DS$ and $FL$ states.  

The changes in the dynamics due to the introduction of repulsive interactions are quantified by the size of the synchronised cluster in that state. Therefore, we examine the variation in the fraction of synchronised oscillators, $f_{sync}$ in chimera states $C_1$, $IPC_2$, and $APC_2$ as $p$ increases. For this, we consider $G(x)$ that gives rise to $IPC_2$ and $APC_2$ at $p$=0 (attractive coupling) and plot $f_{sync}$ as $p$ increases in Fig.~\ref{fig:Fig10}. The subfigures in the top panel show the variation of $f_{sync}$ with $p$ when the repulsive connections are introduced only in the attractive coupling region (keeping the neutral coupling region unchanged). We observe that as the repulsive connections increase, indicated by an increase in $p$, the number of synchronised oscillators in $IPC_2$ first decreases, resulting in a desynchronized state for intermediate values of $p$ and then increases with the emergence of $APC_2$ as repulsive coupling dominates and $p$ approaches 1 (Fig.~\ref{fig:Fig10}(a)). On the contrary, if we begin with $G(x)$ that results in $APC_2$ at $p=0$, the system evolves to a desynchronized state as repulsive coupling, $p$ increases (Fig.~\ref{fig:Fig10}(b)). We compare these results with the scenario when the repulsive coupling is introduced in the entire coupling range (including the zero coupling region) in $G(x)$. For this case, the variation of $f_{sync}$ with $p$ is shown in the lower panel of Fig.~\ref{fig:Fig10}. From the plots, we observe that when the repulsive interactions are introduced in the entire range, $IPC_2$ and $APC_2$ collapse to a desynchronized state (Fig.~\ref{fig:Fig10}(c),(d)). The coupling function $G(x)$, and the corresponding asymptotic phase distribution of the system at $p=0$ and $p=1$ are shown in the insets. Therefore, from Fig.~\ref{fig:Fig10} we conclude that the transition in states with repulsive coupling probability, $p$ depends on the type of coupling kernel and the manner in which repulsive connections are introduced.

We study the collapse of chimeras to a desynchronized state with an increase in repulsive coupling probability $p$ and observe that the collapse occurs at a higher $p$ for $C_1$ compared to the two-cluster chimeras ($IPC_2$ and $APC_2$). Among $IPC_2$ and $APC_2$, the in-phase state, although having a higher $f_{sync}$ at $p=0$, goes to desynchronization at a lower $p$ compared to the antiphase state. Moreover, the collapse from chimera state to the desynchronized state shows power law behavior. A log-log plot of $f_{sync}$ v/s $p$ for transition to desynchronized state from $C_1$, $IPC_2$, and $APC_2$ states is shown in Fig.~\ref{fig:Fig11} and the collapse follows a power law such that $f_{sync}=p^{e}$ where the exponent $-3<e<-1$ for different chimeras.

\section{Effect of repulsive coupling on multi-cluster Chimeras}\label{sec:multi}

\begin{figure}[htbt!]
	\centering
	\includegraphics[width=1.0\linewidth]{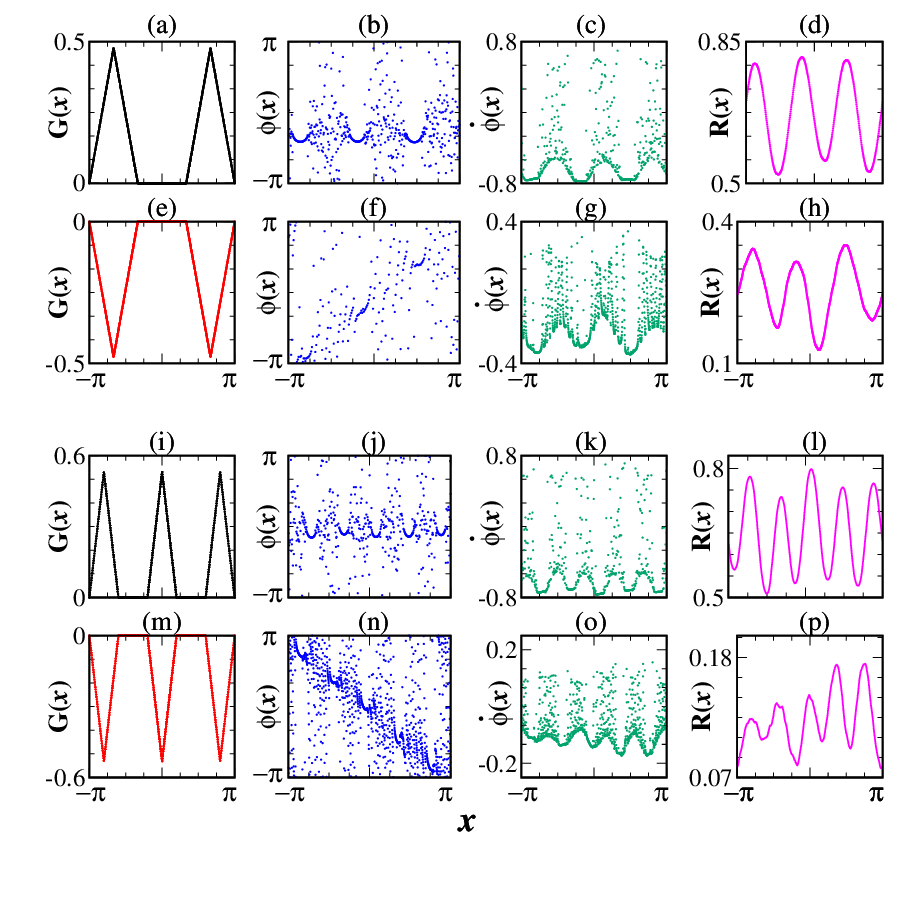}
	\caption{ The first column shows the coupling function $G(x)$, the second column presents the phase snapshots $\phi(x)$ , the third column shows the corresponding $\dot{\phi}(x)$ and the last column shows the order parameter $R(x)$. When the coupling $G(x)$ is attractive, the asymptotic states are shown:  (a)-(d) 3 cluster in-phase chimeras ($IPC_3$) and (i)-(l) 5 cluster in-phase chimeras  ($IPC_5$). When the attractive coupling leading to an odd number of in-phase cluster chimera becomes repulsive (keeping zero coupling region unchanged) the resulting states are demonstrated as: (e)-(h) 3 cluster splay chimera ($SC_3$) and (m)-(p) 5 cluster splay chimera ($SC_5$). The plots are obtained for $N$ = 1024 after discarding 5 $\times10^5$ transients.}
	\label{fig:Fig12}
\end{figure}

Extending the investigation of the two-cluster chimeras, we consider the effects of the introduction of repulsive coupling on multi-cluster chimeras. The multi-cluster chimeras are obtained using the prescription given in our earlier work~\cite{ujjwal2013} for the piecewise linear kernel $G(x)$.  Using the described algorithm chimeras with a desired number of in-phase and antiphase clusters can be obtained. The effect of making the attractive coupling repulsive, keeping the zero coupling region unaltered, varies depending upon whether the number of clusters in the chimera state is odd or even. For a chimera with an odd number of in-phase clusters, the repulsive coupling leads to splay chimeras (Fig.~\ref{fig:Fig12}(f),(n)). The state is characterized by the emergence of equally spaced synchronized clusters with a phase difference of $2\pi /n$ where $n$ represents the number of clusters. In splay chimeras, the consecutive synchronized clusters have equal phase difference and are separated by desynchronized group of oscillators. The splay chimera with $n$ number of clusters is denoted by $SC_n$. The in-phase chimeras with 3 and 5 clusters denoted by $IPC_3$ and $IPC_5$ respectively obtained from piecewise attractive coupling $G(x)$ are shown in Fig.~\ref{fig:Fig12}(b),(j). Their counterpart for the repulsive coupling are terms as $SC_3$ and $SC_5$ are shown in Fig.~\ref{fig:Fig12}(f),(n).

\begin{figure}[htbt!]
	\centering
	\includegraphics[width=1.0\linewidth]{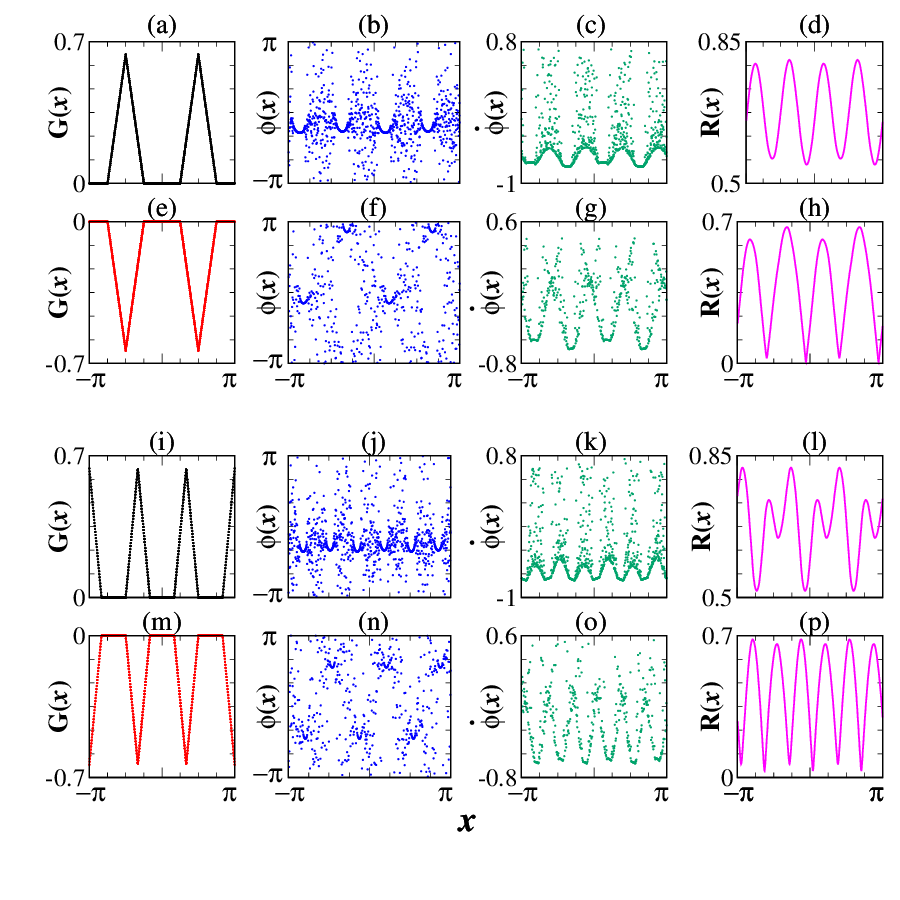}
	\caption{The first column shows the coupling function $G(x)$ and the resulting phase profile $\phi(x)$ is shown in the second column. The third and fourth columns show the frequency $\dot{\phi}(x)$ and order parameter $R(x)$ profiles of the system respectively. When the coupling $G(x)$ is attractive, the asymptotic states are shown:  (a)-(d) 4 cluster in-phase chimeras ($IPC_4$), and (i)-(l) 6 cluster in-phase chimeras ($IPC_6$). When the attractive coupling regions become repulsive (keeping zero coupling regions unchanged) the resulting states are demonstrated as: (e)-(h) 4 cluster antiphase chimeras ($APC_4$), and (m)-(p) 6 cluster antiphase chimera ($APC_6$). }
	\label{fig:Fig13}
\end{figure} 

\pagebreak
\begin{figure}[htbt!]
	
	\includegraphics[width=1.0\linewidth]{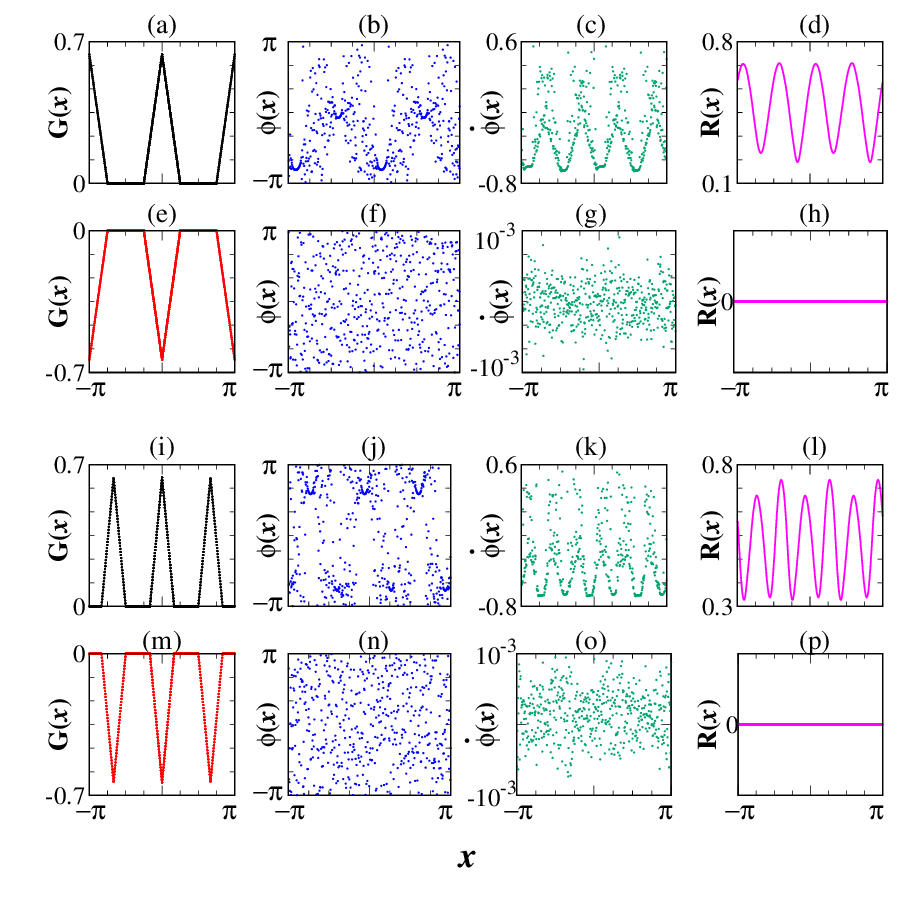}
	\caption{Figure showing emergence of desynchrony from $APC_{n}$ when $n$ is even. The first column shows $G(x)$, the second column presents the phases $\phi(x)$ , the third column shows the corresponding frequency $\dot{\phi}(x)$  and the rightmost column shows $R(x)$. When the coupling $G(x)$ is attractive, the asymptotic states are shown:  (a)-(d) 4 cluster antiphase chimeras($APC_4$), and (i)-(l) 6 cluster antiphase chimeras($APC_6$). When the attractive coupling regions turn repulsive (keeping zero coupling regions unchanged) the resulting state is desynchronized as demonstrated in (e)-(h) and (m)-(p). }
	\label{fig:Fig14}
\end{figure}

An attractive $G(x)$ can result in chimeras with an even number of synchronized clusters where two possibilities arise: all clusters are in-phase or alternate clusters are antiphase to each other. First, we consider $G(x)$ that gives rise to chimeras with an even number of in-phase clusters. When the attractive coupling regions are made repulsive in such $G(x)$, the resulting spatio-temporal pattern is chimeras with antiphase synchronized clusters. Therefore, there is a switching from $n$-cluster in-phase chimera ($IPC_n$) to $n$-cluster antiphase chimera ($APC_n$). This observation is consistent for chimeras with 4, 6 and higher numbers of clusters. For illustration, this switching is shown for 4 and 6 cluster chimeras in Fig.~\ref{fig:Fig13}.  The attractive coupling kernel $G(x)$, which leads to antiphase chimeras with an even number of clusters, gives rise to a fully desynchronized state when the positive (attractive) coupling region becomes negative (repulsive). The results for $G(x)$ resulting in 4 and 6 cluster chimeras are shown in Fig.~\ref{fig:Fig14}.  

From our investigation on the multi-cluster chimeras, we can identify patterns in the emergent states. A coupling function with repulsive local interactions will lead to a desynchronized state in most cases. The interaction scheme that leads to chimeras with an odd number of in-phase clusters can make these clusters splay if the attractive coupling becomes repulsive. Repulsive coupling can also lead to switching of phases from in-phase to antiphase in chimeras with an even number of clusters. Also, the range in which $R(x)$ lies helps us identify different dynamical states of the system. A lower value of $R$ indicates weaker coherence among the oscillators, while a higher value corresponds to stronger coherence. $R(x)$ profiles allow us to compare the states according to their level of coherence. Among the observed states, $IPC_n$ shows higher degree of coherence indicated by a higher value of $R(x)$ ( $0.5<R(x)<0.9$)), followed by $APC_n$ ($0<R(x)<0.8$), while the $SP_n$ shows the lowest coherence  ($0.05<R(x)<0.5$). Attractive coupling generally favours synchronization while repulsive is favourable to desynchrony. For multi-cluster chimera states, we observe repulsive coupling can induce phase lag between the synchronized clusters, hence giving rise to antiphase and splay chimeras that have overall weak coherence as compared to the in-phase states. Therefore, a piecewise linear repulsive coupling can lead to splay states for low $\alpha$ and splay chimeras for higher $\alpha$ values. The resultant states for piecewise linear attractive and repulsive couplings (with zero coupling regions) are summarized in Table~\ref{Table2}.

   \begin{table}[h!]
  	\centering
  	\renewcommand{\arraystretch}{1.8}
  	\begin{tabular}{|c|c|c|c|}
  		\hline
  		& \textbf{Odd $n$} & \multicolumn{2}{c|}{\textbf{Even $n$}} \\ \hline
  		\textbf{Piecewise attractive} & $IPC_n$ & $IPC_n$ & $APC_n$ \\ \hline
  		\textbf{Piecewise repulsive} & $SC_n$ & $APC_n$ & $DS$ \\ \hline
  	\end{tabular}
  	\caption{Table summarizing states for finite-range piecewise linear purely attractive and purely repulsive coupling (with regions of neurtral coupling) where $IPC_n$ stands for chimeras with $n$ numbers of in-phase synchronized clusters, $APC_n$ denotes chimeras with $n$ antiphase clusters, and $SC_n$ is a splay chimera with $n$ clusters. $DS$ is the desynchronized state.}
  	\label{Table2}
  \end{table}

 \section{Ott-Antonsen Reduction}\label{sec:dim-reduction}

  \begin{figure*}[htbt!] 	
   	\includegraphics[width=1\linewidth]{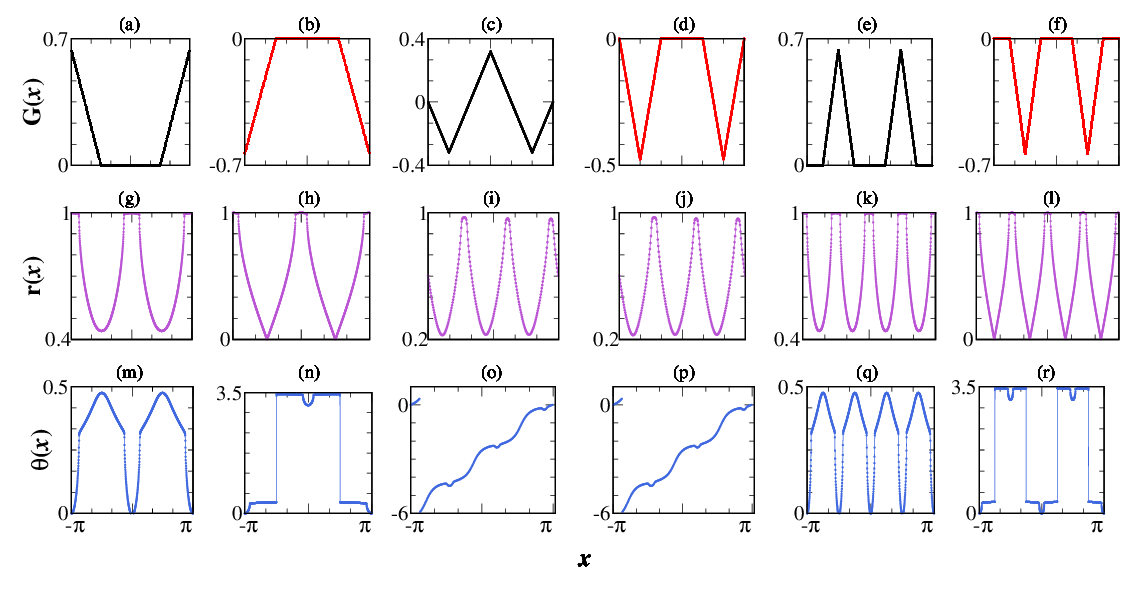}
   	\vspace{-10 pt}
   	\caption{The first row [(a)-(f)] shows the coupling kernel $G(x)$ giving rise to different cluster chimeras. The second [(g)-(l)] and the third [(m)-(r)] rows depict the stable $r(x)$ and $\theta(x)$ profiles computed from Eq.\eqref{eq:eq29} - Eq.\eqref{eq:eq30} for the respective $G(x)$. $G(x)$, $r(x)$, and $\theta(x)$ profiles are shown for $IPC_2$ in first column [(a),(g),(m)]; $APC_2$ in second coulmn [(b),(h),(n)] ; $SC_3$ in third [(c),(i),(o)] and fourth columns [(d),(j),(p)]; $IPC_4$ in fifth column [(e),(k),(q)] and the last column [(f),(l),(r)] is for $APC_4$. The parameters used are: $\gamma$ = -0.74051, $D$ = 0.08, $\beta$ = 0.05 and $N$ = 1024.}
   	\label{fig:Fig15}
   \end{figure*}

 The Ott and Antonsen(OA) analysis~\cite{Ott2008,Ott2009} can be employed to reduce the $N$-dimensional system to a finite set of equations, in the limit of large $N$. Eq.\eqref{eq:eq1} can be rewritten as,
 \begin{equation}\label{eq:eq14}
 	\frac{\partial \phi(x,t)}{\partial t} = \omega(x) - \int_{-\pi}^{\pi}G(x-x')\cos(\phi(x,t) - \phi(x',t) - \beta)dx',
 \end{equation}

 where the natural frequencies $\omega$'s are taken from the distribution $g(\omega)$ and $\beta$ = $\pi/2 - \alpha$. 
 
In the continuum limit ($N \to \infty$), the state of the system can be characterised by the probability density function $f(x, \omega, \phi, t)$. This function satisfies the continuity equation since the number of oscillators is conserved.
 
 \begin{equation}\label{eq:eq15}
 	\frac{\partial f}{\partial t} + 	\frac{\partial }{\partial \phi}(fv)= 0,
 \end{equation} 
 where $v$ is the velocity of the oscillators given by

	\begin{equation}
		\small
		v=\omega-\int_{-\pi}^{\pi}\int_{-\infty}^{\infty}\int_{-\pi}^{\pi}
		G(x-x')\cos(\phi-\phi'-\beta)f(x',\omega,\phi',t)\,d\phi' d\omega dx'.
	\end{equation}

Here we use the shorthand notation such that $\phi$ = $\phi(x)$ and $\phi'$ = $\phi(x')$. We can write the order parameter in terms of the probability density function, $f(x, \omega, \phi, t)$ as

  \begin{equation}\label{eq:eq17}
  	R(x,t) \equiv \int_{-\pi}^{\pi} \int_{-\infty}^{\infty}\int_{-\pi}^{\pi}G(x-x')e^{-i\phi'}f(x', \omega, \phi', t)d\phi'd\omega dx'.
  \end{equation}
 \normalsize
  
  Rewriting $v$ in terms of $R$, we get
  
  \begin{equation}\label{eq:eq18}
  	v = \omega - \frac{1}{2}[Re^{i(\phi - \beta)}+\bar{R}e^{-i(\phi - \beta)}]
  \end{equation}
where an overbar represents the complex conjugate.
 
 Now, expanding $f(x', \omega,\phi',t)$ in a Fourier series,
 
 \begin{equation}\label{eq:eq19}
 	f(x',\omega,\phi',t) = \frac{g(\omega)}{2\pi}\left[ 1 + 
 	\begin{cases}
 	\sum_{n=1}^{\infty}k_n(x',\omega,t)e^{in\phi'}+ c.c
 	\end{cases}
     \right] 
 \end{equation}

 where c.c indicates the complex conjugate of the previous term. Taking $k_{n}(x',\omega,t)= [q(x',\omega,t)]^{n}$. Substituting the expressions of $f$ and $v$ in Eq.\eqref{eq:eq15} we get,
   \begin{equation}\label{eq:eq20}
  	\frac{\partial q}{\partial t} = -i\omega q + \frac{i}{2}\left[ Re^{-i\beta} + \bar{R}e^{i\beta}q^{2}\right].
  \end{equation}

 The order parameter can be written as,
  
  \begin{equation}\label{eq:eq21}
  	R(x,t) = \int_{-\pi}^{\pi}\int_{-\infty}^{\infty} G(x-x')g(\omega)q(x',\omega,t)d\omega dx'.
  \end{equation}
 
 Assuming that $\omega$ is taken from a Lorentzian distribution with half-width-at-half-maxima and centre of the distribution as $D$ and $\omega_o$ respectively such that,
  
 \begin{equation}\label{eq:eq22}
 	g(\omega) = \frac{D/\pi}{(\omega - \omega_o)^2 + D^{2}}.
 \end{equation}
  We set $\omega_o$=0 by moving to a rotating frame. Now, performing contour integration over $\omega$ in Eq.\eqref{eq:eq21} we get,
  
  \begin{equation}\label{eq:eq23}
  	R(x,t) = \int_{-\pi}^{\pi}G(x-x')q(x',-iD,t) dx'.
  \end{equation}
 
 Let $\hat{z}(x,t) = q(x,-iD,t)$, then Eq.\eqref{eq:eq20} becomes,
    
     \begin{equation}\label{eq:eq24}
    	\frac{\partial \hat{z}}{\partial t} = -D\hat{z} + \frac{i}{2}\left[ Re^{-i\beta} + \bar{R}e^{i\beta}\hat{z}^{2}\right]
    \end{equation}
   where
     \begin{equation}\label{eq:eq25}
     	R(x,t) = \int_{-\pi}^{\pi}G(x-x')\hat{z}(x',t) dx'.
     \end{equation}
   
Further simplification can be done by transforming to a rotating coordinate frame. Thus, we write $z \equiv \hat{z} e^{i\gamma t}$ where $\gamma$ is the angular frequency of rotation. We find that $z$ satisfies 
    
    \begin{equation}\label{eq:eq26}
    	\frac{\partial z}{\partial t} = (i\gamma-D)z + \frac{i}{2}\left[ \hat{R}e^{-i\beta} + \bar{\hat{R}}e^{i\beta}z^{2}\right]
    \end{equation}
    where
    \begin{equation}\label{eq:eq27}
    	\hat{R}(x,t) = \int_{-\pi}^{\pi}G(x-x')z(x',t) dx'.
    \end{equation}
   
   The complex variable $z$ can be written in polar coordinates ($r$,$\theta$) as,
   \begin{equation}\label{eq:eq28}
   	z(x) = r(x)e^{-i\theta(x)}
   \end{equation}
   Using $r$ = $r(x)$ and $r'$ = $r(x')$,
   the evolution equations for the coordinates $r$ and $\theta$ for each oscillator will be,
   \begin{equation}\label{eq:eq29}
   	\frac{ dr}{dt} = -Dr+\frac{(1-r^2)}{2} \int_{-\pi}^{\pi}G(x-x')r' \sin(\theta' - \theta + \beta) dx',
   \end{equation}
   
   \begin{equation}\label{eq:eq30}
   	\frac{ d\theta}{dt} = -\gamma-\frac{(1+r^2)}{2r} \int_{-\pi}^{\pi}G(x-x')r' \cos(\theta' - \theta + \beta)dx'.
   \end{equation}

Eqs.\eqref{eq:eq29}-\eqref{eq:eq30} are solved with $\gamma$ estimated from the stationary solution of Eq. \eqref{eq:eq26} [See \cite{laing2009}]. The obtained $r(x)$ and $\theta(x)$ profiles predict the qualitative dynamics of the chimera state for a given $G(x)$ correctly. The number of peaks in $r(x)$ corresponds to the number of synchronized clusters, while the phase difference between the clusters can be deduced from the $\theta(x)$ profile for a particular chimera state (Fig. \ref{fig:Fig15}). However, quantitatively there could be some mismatch between the $R(x)$ profiles obtained numerically from Eqs.\eqref{eq:eq1}-\eqref{eq:eq2} and the ones obtained from dimensional reduction (Eqs.\eqref{eq:eq29}-\eqref{eq:eq30}) due to the non-zero value of $D$ considered therein. From Fig. \ref{fig:Fig15}, we note that the Ott-Antonsen technique can detect emergent states in the system and also distinguishes between the in-phase and antiphase clustered chimera state. This analysis confirms the formation of different multi-cluster chimeras and phase switching when attractive coupling becomes repulsive.

 \section{Summary}\label{sec:summary}

Here we have investigated the dynamical states that arise due to repulsive nonlocal coupling in an ensemble of 
Sakaguchi-Kuramoto oscillators on a ring. The couplings we have considered are all piecewise linear and symmetric;
this complements our earlier work wherein attractive nonlocal coupling had been considered. 

Phase frustration, along with the nonlocal interaction, helps to create the multi-cluster chimeras. We observe that 
repulsive interactions induce a phase shift between the synchronized oscillators, giving rise to splay states 
for low phase lag parameter $\alpha$, while for large $\alpha$ we obtain splay chimeras or chimeras with antiphase clusters. 
In the splay chimera state, the desynchronized groups of oscillators coexist with $n$ synchronized clusters, there being
a phase difference of $2\pi/n$ between consecutive clusters.  

The effects of attractive and repulsive coupling can be contrasted in a simple manner. 
Consider a multi-cluster chimera formed with attractive piecewise-linear coupling, and flip the signs so that the 
attractive regions become repulsive. We  observe a phase shift between the synchronized clusters, depending on the 
number of clusters. If there had been a chimera with an odd number of in-phase clusters, this becomes a splay chimera,
whereas a chimera with an even number of clusters changes into a one with antiphase clusters. 
Our investigation suggests that repulsive coupling favors splay chimeras for high values of the phase-lag. The results 
are robust with respect to initial conditions and change in the form of the coupling function.

Phase shifts are significant and often employed for signal modulation and optimizing transmission stability in the 
field of telecommunication, signal processing, and electronics~\cite{gharsalli2026}. In neural networks, phase shifts 
in firing patterns of coupled neurons are linked to sensory processes such as communication~\cite{tiesinga2010}, 
spatial navigation~\cite{qasim2021} and also some pathological conditions~\cite{sedghizadeh2022}.  Several mechanisms 
have been proposed to control the phase difference between the dynamics of coupled units, for example, in neuronal 
networks, the phase shift between the firing of neurons can be modulated by applying an external stimulus~\cite{stigen2011}. 
Our study proposes that this control of phase differences among synchronised group of oscillators can also be 
achieved by introducing repulsive or inhibitory connectivity in the system. The results presented in this work are 
interesting and open new directions for exploration. It has been found that new kinds of chimeric patterns can arise in 
modular networks~\cite {ujjwal2016}. It would be interesting to study the interplay between distance-dependent attractive 
and repulsive couplings in hierarchical networks. Also, the possibility of swarm chimeras and multicoherent phase chimera 
states~\cite{cai2024} can be investigated in the generalized Kuramoto model by considering the higher harmonic modes
in the interaction term. 
  
\begin{acknowledgments}
AS and SRU acknowledge the resources provided by PARAM Shivay computational facility at the Indian Institute of Technology, 
Varanasi. SRU is thankful to Banaras Hindu University for financial support under the Institute of Eminence (IoE) scheme.  
\end{acknowledgments}



\begin{thebibliography}{47}

\bibitem{majhi2020}S. Majhi, S. N. Chowdhury, and D. Ghosh, 
\href{https://iopscience.iop.org/article/10.1209/0295-5075/132/20001}{\epl{132}{20001}{2020}}.

\bibitem{Neocort2017} R. Tatti, M. S. Haley, O. K. Swanson, T. Tselha, and A. Maffei, 
\href{https://doi.org/10.1016/j.biopsych.2016.09.017}{\biopsych{81}{821--831}{2017}}.


\bibitem{ermentrout} B. Ermentrout, 
\href{https://doi.org/10.1088/0034-4885/61/4/002}{\rpp{61}{353}{1998}}.

\bibitem{Schwartz} E. L. Schwartz, {\sl Computational neuroscience} (MIT Press, 1993).

\bibitem{hong2011} H. Hong and S. H. Strogatz,
\href{https://doi.org/10.1103/PhysRevLett.106.054102}{\prl{106}{054102}{2011}}.

\bibitem{Giron2016} A. Gir{\'o}n, H. Saiz, F. S. Bacelar, Roberto F. S. Andrade, and J. G{\'o}mez-Gardeñes, 
\href{https://doi.org/10.1063/1.4952960}{\ch{26}{065302}{2016}}.

\bibitem{paul2024} B. Paul, B. Bandyopadhyay, and T. Banerjee, 
\href{https://doi.org/10.1103/PhysRevE.110.034210}{\pre{110}{034210}{2024}}.

\bibitem{strogatz1993} S. H. Strogatz and I. Stewart,
\href{https://doi.org/10.1038/scientificamerican1293-102}{\sciam{269}{102}{1993}}.

\bibitem{strogatz1993splay} S. H. Strogatz and R. E. Mirollo, \href{https://journals.aps.org/pre/abstract/10.1103/PhysRevE.47.220}{\pre{47}{220}{1993}}.

\bibitem{hadley1987} P. Hadley and M. R. Beasley,
\href{https://doi.org/10.1063/1.98100}{\apl{50}{621--623}{1987}}.

\bibitem{kur2002} Y. Kuramoto and D. Battogtokh, 
\href{http://www.j-npcs.org/abstracts/vol2002/v5no4/v5no4p380.html}{\npcs{5}{380--385}{2002}}.

\bibitem{abrams2004} D. M. Abrams and S. H. Strogatz, 
\href{https://doi.org/10.1103/PhysRevLett.93.174102}{\prl{93}{174102}{2004}}.
	

\bibitem{panaggio2015} M. J. Panaggio and D. M. Abrams, 
\href{https://doi.org/10.1088/0951-7715/28/3/R67}{\nonlin{28}{R67}{2015}}.


\bibitem{rattenborg} N. C. Rattenborg, C. J. Amlaner, and S. L. Lima, 
\href{https://doi.org/10.1016/S0149-7634(00)00039-7}{\nbr{24}{817}{2000}}.


\bibitem{haugland2021} S. W. Haugland, 
\href{https://doi.org/10.1088/2632-072X/ac0810}{\jpcx{2}{032001}{2021}}.

\bibitem{mathews2006} C. G. Mathews, J. A. Lesku, S. L. Lima, and C. J. Amlaner, 
\href{https://doi.org/10.1111/j.1439-0310.2006.01138.x}{\eth{112}{286}{2006}}.

\bibitem{majhi2019} S. Majhi, B. K. Bera, D. Ghosh, and M. Perc,  
\href{https://doi.org/10.1016/j.plrev.2018.09.003}{\plr{28}{100}{2019}}.


\bibitem{bansal2019} K. Bansal, J. O. Garcia, S. H. Tompson, T. Verstynen, {\em et al.}, 
\href{https://doi.org/10.1126/sciadv.aau8535}{\sciadv{5}{eaau8535}{2019}}.

\bibitem{makinwa2023} T. Makinwa, K. Inaba, T. Inagaki, Y. Yamada 
{\em et al.},
\href{https://doi.org/10.1038/s42005-023-01240-x}{\comphys{6}{121}{2023}}.


\bibitem{cherry2008} E. M. Cherry and F. H. Fenton, 
\href{https://doi.org/10.1088/1367-2630/10/12/125016}{\njp{10}{125016}{2008}}.


\bibitem{martens2013} E. A. Martens, S. Thutupalli, A. Fourri\`ere, and O. Hallatschek, 
\href{https://doi.org/10.1073/pnas.1302880110}{\pnas{110}{10563}{2013}}.

\bibitem{ebrah2022} P. Ebrahimzadeh, M. Schiek, and Y. Maistrenko,  
\href{https://doi.org/10.1063/5.0103071}{\ch{32}{103118}{2022}}.


\bibitem{tinsley2012} M. R. Tinsley, S. Nkomo, and K. Showalter, 
\href{https://doi.org/10.1038/nphys2371}{\natphys{8}{662--665}{2012}}.

\bibitem{matheny2019} M. H. Matheny, J. Emenheiser, W. Fon, A. Chapman {\em et al.}, 
\href{https://doi.org/10.1126/science.aav7932}{\sc{363}{eaav7932}{2019}}.

\bibitem{avella2014} J. C. Gonz{\'a}lez-Avella, M. G. Cosenza, and M. San Miguel, 
\href{https://doi.org/10.1016/j.physa.2013.12.035}{\physa{399}{24}{2014}}.




\bibitem{omel2012} I. Omelchenko, O. E. Omel'chenko, P. H\"ovel, and E. Sch\"oll, 
\href{https://doi.org/10.1103/PhysRevLett.110.224101}{\prl{110}{224101}{2013}}.

\bibitem{ujjwal2013} S. R. Ujjwal and R. Ramaswamy, 
\href{https://doi.org/10.1103/PhysRevE.88.032902}{\pre{88}{032902}{2013}}.
 
\bibitem{sathiyadevi2018} K. Sathiyadevi, V. K. Chandrasekar, D. V. Senthilkumar, and M. Lakshmanan, 
\href{https://doi.org/10.1103/PhysRevE.97.032207}{\pre{97}{032207}{2018}}.

\bibitem{ermentrout1994} G. B. Ermentrout and N. Kopell, 
\href{https://doi.org/10.1137/S0036139992231964}{\sjam{54}{478}{1994}}.


\bibitem{Hutt2014} M.-T. H{\"u}tt, M. Kaiser, and C. C. Hilgetag, 
\href{https://doi.org/10.1098/rstb.2013.0522}{\ptrsb{369}{20130522}{2014}}.

\bibitem{sakaguchi1986}H. Sakaguchi and Y. Kuramoto,
\href {https://academic.oup.com/ptp/article/76/3/576/1875056}{\prtp{76}{576--581}{1986}}.

\bibitem{hutt2003} A. Hutt, M. Bestehorn, and T. Wennekers, 
\href{https://doi.org/10.1088/0954-898X/14/2/310}{\ncns{14}{351}{2003}}.


\bibitem{zillmer2006} R. Zillmer, R. Livi, A. Politi, and A. Torcini, 
\href{https://doi.org/10.1103/PhysRevE.74.036203}{\pre{74}{036203}{2006}}.


\bibitem{Jordan} D. W. Jordan and P. Smith,  Nonlinear Ordinary Differential Equations, \href{http://ndl.ethernet.edu.et/bitstream/123456789/38112/1/41.pdf}{Oxford University Press, (2007)}.


\bibitem{Ott2008} E. Ott and T. M. Antonsen, 
\href{https://doi.org/10.1063/1.2930766}{\ch{18}{037113}{2008}}.

\bibitem{Ott2009} E. Ott and T. M. Antonsen, 
\href{https://doi.org/10.1063/1.3136851}{Chaos {\bf 19}, 023117 (2009)}.

\bibitem{laing2009} C. R. Laing, 
\href{https://doi.org/10.1016/j.physd.2009.04.012}{\physd{238}{1596}{2009}}.

\bibitem{gharsalli2026} S. Gharsalli, R. Aloui, S. Mhatli, A. Corona-Chavez {\em et al.}, 
\href{https://www.jpier.org/PIERB/pier.php?paper=25082704}{\pierb{116}{33--47}{2026}}.

\bibitem{tiesinga2010} P. H. Tiesinga and T. J. Sejnowski, 
\href{https://www.frontiersin.org/journals/human-neuroscience/articles/10.3389/fnhum.2010.00196/full} {\fhn{4}{196}{2010}}. 

\bibitem{qasim2021}S. E. Qasim, I. Fried, and J. Jacobs, 
\href{https://www.sciencedirect.com/science/article/pii/S0092867421004967?}{\cell{184}{3242}{2021}}.

\bibitem{sedghizadeh2022} M. J. Sedghizadeh, H. Aghajan, Z. Vahabi, S. N. Fatemi, and A. Afzal, 
\href{https://link.springer.com/article/10.1007/s00429-022-02554-2}{\bsf{227}{2957}{2022}}.

\bibitem{stigen2011} T. Stigen, P. Danzl, J. Moehlis, and T. Netoff, 
\href{https://journals.physiology.org/doi/full/10.1152/jn.00898.2011} {\jneurophysiol{105}{2074}{2011}}.

\bibitem{ujjwal2016} S. R. Ujjwal, N. Punetha, and R. Ramaswamy, 
\href{https://journals.aps.org/pre/pdf/10.1103/PhysRevE.93.012207} {\pre{93}{012207}{2016}}.

\bibitem{cai2024} Z. Cai, Z. Liu, S. Guan, J. Kurths, and Y. Zou, 
\href{https://doi.org/10.1103/PhysRevLett.133.227201}{\prl{133}{227201}{2024}}.

\end{thebibliography}
\end{document}